%% file: cbc_frb_driver.tex
\documentclass[fleqn,usenatbib]{mnras}
\usepackage{amsmath,color,bm}
\usepackage{txfonts}
\usepackage{multirow}
\usepackage{epsfig}

\usepackage{booktabs}
\usepackage{subfig}

\makeatletter
\renewcommand{\p@subfigure}{}
\makeatother
\usepackage{nameref}
\usepackage{graphicx}
\usepackage{graphics}
\usepackage[space]{grffile}
\usepackage{dblfloatfix}
\usepackage{url}
\usepackage[utf8]{inputenc}
\usepackage{fancyref}
\usepackage{xcolor}
\usepackage[normalem]{ulem}
\usepackage{eqnarray,amsmath}
\usepackage{multicol}

\definecolor{magenta}{rgb}{139, 0, 139}

\input{title.tex}

\begin{document}
\label{firstpage}
\pagerange{\pageref{firstpage}--\pageref{lastpage}}
\maketitle

\input{abstract.tex}

\section{Introduction}\label{sec:introduction}
	\input{introduction.tex}

\section{Motivation and Method}\label{sec:method}
\input{method.tex}

\section{Results}\label{sec:results}
	\input{results.tex}

\section{Summary and Outlook}\label{sec:conclusion}
	\input{conclusion.tex}

\section*{Acknowledgements}

We are grateful to Jose Mar\'ia Ezquiaga for carefully reviewing our manuscript and providing useful comments. We also thank Souvik Jana for helpful discussions. MKS's and PA's research was supported by the Department of Atomic Energy, Government of India, under project no. RTI4001. PA's research was funded by the Max Planck Society through a Max Planck Partner Group at ICTS-TIFR and by the Canadian Institute for Advanced Research through the CIFAR Azrieli Global Scholars program. SPT is a CIFAR Azrieli Global Scholar in the Gravity and Extreme Universe Program. Numerical calculations reported in this paper were performed using the Alice cluster at ICTS-TIFR.

\section*{Data Availability}
The data underlying this article will be shared on reasonable request to the corresponding author.

\bibliography{references}
\newpage

\onecolumn
\section*{}\label{sec:appendix}
	\input{appendix.tex}

\end{document}

%% file: title.tex
\title[CBC-FRB Association with Lensing]{Associating fast radio bursts with compact binary mergers via gravitational lensing}

\author[M. K. Singh et al.]{Mukesh Kumar Singh,$^1$\thanks{E-mail: mukesh.singh@icts.res.in}
  Shasvath J. Kapadia,$^{1, 2}$
  Soummyadip Basak, $^1$
  Parameswaran Ajith, $^{1,5}$ \newauthor
  and
  Shriharsh P. Tendulkar,$^{3, 4, 5}$ \\
  $^1$International Centre for Theoretical Sciences, Tata Institute of Fundamental Research, Bangalore 560089, India\\
  $^2$Inter University Centre for Astronomy and Astrophysics, Post Bag 4, Ganeshkhind, Pune - 411007, India\\
  $^3$Department of Astronomy and Astrophysics, Tata Institute of Fundamental Research, Mumbai - 400005, India \\
  $^4$National Centre for Radio Astrophysics, Post Bag 3, Ganeshkhind, Pune - 411007, India \\
  $^5$Canadian Institute for Advanced Research, CIFAR Azrieli Global Scholar, MaRS Centre, West Tower, 661 University Ave, Toronto, ON M5G 1M1, Canada}

%% file: abstract.tex
\begin{abstract}
The origin of fast radio bursts (FRBs) is currently an open question with several proposed sources and corresponding mechanisms for their production. Among them are compact binary coalescences (CBCs) that also generate gravitational waves (GWs). Spatial and temporal coincidences between GWs and FRBs have so far been used to search for potential FRB counterparts to GWs from CBCs. However, such methods suffer from relatively poor sky-localisation of the GW sources, and similarly poor luminosity distance estimates of both GW and FRB sources. The expected time delay between the GW and radio emission is also poorly understood. In this work, we propose an astrophysical scenario that could potentially provide an unambiguous association between CBCs and FRBs, if one exists, or unambiguously rule out FRB counterparts to a given CBC GW event. We demonstrate that, if a CBC that emitted both GWs and FRBs, is gravitationally lensed, we can make a $> 5\sigma$ association using time-delay estimates of the lensed GW and FRB images (in strong lensing), which are expected to be measured with mili-second (for GW) and nano-second (FRB) precisions. We also demonstrate that the CBC-FRB association can be made in the microlensing regime as well where wave-optics effects modulate the GW waveform. We further investigate the rate of such detected associations in future observing scenarios of both GW and radio detectors.
\end{abstract}

\begin{keywords}
gravitational waves, fast radio bursts, gravitational lensing:strong, gravitational lensing:micro
\end{keywords}

%% file: introduction.tex
Fast Radio Bursts (FRBs) are short-duration radio bursts whose provenance has still to be ascertained. Since their first detection in 2007 \citep{lorimer_FRB}, there have been $\mathcal{O}(600)$ detections so far \citep{frb_catalog_thornton_et_al, frb_arecibo, frb_masui_et_al, frbs_champion_et_al, frb_catalog_petroff_et_al, frb_caleb_et_al, frb_bannister_et_al, frb_shannon_et_al, CHIME_frb_catalog_old,CHIME_frb_catalog1}. It has been speculated that FRBs can be produced by multiple astrophysical mechanisms \citep{frb_theories_living}. These include magnetospheric emission in rotating neutron stars \citep{frb_BNS_zhang}, core collapse supernovae resulting in the production of a young magnetar \citep{frb_magnetar1, frb_magnetar2}, and compact binary coalescence (CBC) events such as mergers of binary neutron stars (BNSs) via magnetic braking and magnetic reconnection~\footnote{Recently, \cite{FRB_BNS_association} has claimed the association of an FRB signal, FRB 20190425A \citep{CHIME_frb_catalog1}, with a BNS merger, GW190425 \citep{BNS2_LIGO}, at $2.8\sigma$ confidence. But \cite{off_axis_BNS_on_axis_FRB} argue that GW190425 data excludes an on-axis configuration contrary to a small FRB viewing angle for it to be detectable.} \citep{ frb_BNS_merger, frb_BNS_inspiral, frb_review_zhang}, neutron-star black-hole binaries (NSBHs) \citep{frb_charged_BHNS}, and binary black holes (BBHs) \citep{frb_charged_BH} through Poynting flux if at least one of the BHs/NSs in a binary is charged. NSBH mergers could produce a clean FRB-like emission since NS, a spinning magnet, is always charged \citep{frb_charged_BHNS}. If a BNS merger results into a hypermassive NS which is massive enough to collapse into a BH, the magnetic field lines will snap through the BH horizon that can lead to an outward shock of intense short radio burst \citep{frb_supramassive_NS1, frb_supramassive_NS2}. Some models suggest that BNS mergers could also produce coherent radio emission through the excitation of circumbinary plasma by GWs \citep{frb_BNS_circumbinary_plasma}, from the magnetic field that is dynamically generated in the postmerger \citep{frb_precursor_BNS} or from the collision of a GRB forward shock with the surrounding medium \citep{Usov_GRB_shock, Sagiv_GRB_shock}.

On the other hand, gravitational waves (GWs) from $\sim 100$ CBCs have been detected during LIGO-Virgo's first three observing runs \citep{gwtc-1, gwtc-2, gwtc-3, ias-1, ias-2, ias-3, ias-4, Nitz_OGC1, Nitz_OGC2, Nitz_OGC4, Nitz_OGC3}. The majority of these sources are stellar-mass BBHs, although BNSs \citep{gw170817, gw190425} and NSBHs \citep{LVK_NSBHs} have also been observed. GW observations could thus be a means by which CBCs could be associated with FRBs, provided their spatial (sky location and distance) and temporal (time of arrival) parameters could be measured with sufficient precision. However, current GW detectors provide poor sky-volume localisation estimates of CBC sources \citep{Fairhurst_2009, Singer_2014, Singer_2016, distance_uncertainties_GW}. Similarly, the radio observations of FRBs can only weakly constrain their luminosity distance \citep{distance_DM_frb, DM_Xu_Zhang, DM_uncertainties_Xu_et_al, z_DM_distribution_frbs, LVK_CBC_FRB, distance_uncertainties_FRB}. Furthermore, the time of occurrence of a CBC-driven FRB relative to the merger remains an open question. Certain models predict FRBs as precursors to the merger, for example, magnetospheric interaction of two NSs could emit FRBs centuries to decades prior to the merger \citep{frb_BNS_zhang}, while others predict their occurrence postmerger \citep{frb_charged_BH, frb_charged_BHNS, frb_BNS_circumbinary_plasma, frb_precursor_BNS}. Even among those that predict post-merger FRBs, the time delay between GW and radio emission can differ from model to model by several orders of magnitude ($\sim 0 - 5 \times 10^4$s) \citep{NS_collapse_times}. 

It is therefore not surprising that current attempts to associate FRBs with CBCs via temporal and spatial coincidence cannot rule out certain candidate CBC-FRB pairs, while being unable to make the association unambiguously \citep{LVK_CBC_FRB}. In this work, we propose an astrophysical scenario that could potentially enable a confident CBC-FRB association, if one exists. The method crucially relies on the time-delays of images produced by gravitational lensing.

Gravitational lensing of waves that follow null geodesics (like GWs or electromagnetic waves) occurs when they encounter massive objects \citep{lensing_gw_1, lensing_gw_2, lensing_gw_3, lensing_gw_4, lensing_review}. These could either be galaxies or clusters on the one hand, or isolated black holes on the other. If a CBC-FRB pair was lensed by the former, multiple temporally resolvable copies of the GW \citep{Dai_Teja_strong_lensing, Li_et_al_strong_lensing, Ng_et_al_strong_lensing, Oguri_et_al_strong_lensing, Jose_et_al_strong_lensing} and FRB signals \citep{Li_and_Li_frb_lensing, Munoz_et_al_frb_lensing, Dai_and_Lu_frb_lensing, microlensed_frbs_zarif} (with amplitudes rescaled) would be observed. This phenomenon is termed as strong lensing. In particular, the time-delay between the GW images would exactly match the corresponding FRB image time-delay. 

On the other hand, if such a pair was lensed by an isolated black hole (BH) whose gravitational radius is comparable to the GW wavelength, then the GWs and FRBs would be lensed differently. While there would be multiple temporally resolvable FRBs (strong lensing) \citep{Munoz_et_al_frb_lensing,  lensing_frb_review, microlensed_frbs_zarif}, there would be a single GW image whose morphology would be deviated with respect to the unlensed GW (microlensing, due to wave optics effects) \citep{lensing_gw_2, microlensing, Dai_et_al_microlensing, Cheung_et_al_microlensing, Mishra_et_al_microlensing}. Nevertheless, this waveform distortion contains an imprint of the time delay between lensed FRB images that can be extracted assuming a lens model. Even here, the CBC and FRB image time delays must match exactly, if they both were produced by the same source and lens. Previously, \cite{GW_speed_with_lensing1} and \cite{GW_speed_lensing2} have explored the idea of using lensed GW and EM signals to reduce the errors on the measurement of the relative speed of these two signals.

 We show that, given a lensed CBC and a lensed FRB, we can associate the two (if they are produced by the same source and lens) with $> 5\sigma$ confidence. We further estimate the rate of (potential) CBC-FRB associations (if all CBCs containing at least one NS produce FRBs), accounting for the finite field of view of radio telescopes we consider: CHIME \citep{amiri2018chime, CHIME, CHIME_frb_catalog1} and BURSTT \citep{BURSTT}. We find that the rate of NSBH-FRB associations spans $\sim 2 \times 10^{-4} - 3 \times 10^{-3} (4 \times 10^{-4} - 10^{-2}) [2 \times 10^{-3} - 5 \times 10^{-2}] \ \rm{yr}^{-1}$ in O5 (Voyager) [3G] observing scenarios. The association-rates increase to $\sim 0.02 - 0.5 (0.2 - 3) [5 - 87] \ \rm{yr}^{-1}$ when considering the larger field of view of BURSTT. 
The relatively low association-rates are primarily because the radio telescopes do not have an all-sky field of view. Wide field radio telescopes covering a significant fraction of the sky would be highly valuable for this enterprise. Assuming an array of radio telescopes with sensitivities similar to Square Kilometer Array (SKA) \citep{SKA} that can cover the whole sky at a given time, the association rates become as large as $\sim \mathcal{O}(30)\ \left[\mathcal{O}(1800)\right] \ \rm{yr}^{-1}$ for NSBHs and $\sim \mathcal{O}(100)\ \left[\mathcal{O}(8500)\right] \ \rm{yr}^{-1}$ for BNSs in Voyager[3G]. We also compute the BNS-FRB association rates, which are marginally better than NSBH-FRB associations due to the fact that local merger rate of BNSs is higher than NSBHs.

We also provide an example of an NSBH-FRB association where the GW signal from the NSBH is microlensed by a BH, with temporally unresolvable images, but the FRB images are temporally resolved. The modulated waveform can be used to estimate the lensing time delay for the FRB images, and can be compared with the radio observation, to ascertain or rule out the GW-FRB association. We do not quote the rate of such associations because the mass spectrum of microlenses is rather uncertain. 

The rest of the paper is organized as follows. In section \ref{sec:lensing}, we delineate the lensing time-delay equations in both the geometric-optics and wave-optics regimes. We also outline the method used to compute the CBC-FRB association rates given lensing time-delays distributions. We further describe how we assign a false-association-probability (FAP) to assess the association-significance of a lensed CBC-FRB candidate. Section \ref{sec:results} provides illustrative results, and the expected rate of CBC-FRB associations. Section \ref{sec:conclusion} summarizes the work and discusses future prospects of making confident CBC-FRB associations. 

%% file: method.tex
\label{sec:lensing}
\subsection{Gravitational Lensing}
Gravitational lensing is the phenomenon wherein electromagnetic/gravitational waves are deflected from their original path due to mass-inhomogeneities encountered during their propagation \citep{lensing_gw_1, lensing_gw_2, lensing_gw_3, lensing_gw_4}. Given the lensing potential and  the location of the source with respect to the lens, the amount of deflection can be predicted by general relativity (GR) \citep{lensing_review}. Depending on the wavelength of the radiation ($\lambda$) and the gravitational radius ($R_s$) of the lens encountered, gravitational lensing can happen in geometric optics regime ($\lambda \ll R_s$) producing temporally resolvable images, or wave optics regime ($\lambda \sim R_s$) producing additional deformation on the original waveform. We further adopt the thin lens approximation, which is valid for all lensing scenarios where the separation scales between lens, source and observer are cosmological.

\subsubsection{Estimating time-delay from temporally resolvable images}
Consider a lensing scenario where the angular diameter distances between source and observer, lens and observer, lens and source, are given by  $D_s$, $D_l$, and $D_{ls}$, respectively. We further assume the geometric optics limit ($\lambda \ll R_s$), valid for galaxies (or clusters) lensing GWs from stellar-mass CBCs, and FRBs. Let $\textbf{x}$ be the angular coordinates of the images in the lens plane, and $\textbf{y}$ the source coordinates in the source plane. The Fermat potential $\phi(\textbf{x}, \textbf{y})$, evaluated at an image location $\textbf{x}$, can be computed from the geometric separation of the source and image, and the mass-density profile of the lens. The time delay between two images labelled as $1$ and $2$ is then proportional to difference in the Fermat potentials evaluated at those image locations \citep{kormann_et_al}: 
\begin{eqnarray}
    \Delta t_{12} = \frac{(1 + z_l)}{c} \frac{D_s}{D_{l} D_{ls}} \left[ \phi (\textbf{x}_2, \textbf{y}) - \phi (\textbf{x}_1, \textbf{y}) \right]
\end{eqnarray}
where $z_l$ is the lens redshift. In this work, we assume the singular isothermal ellipsoid (SIE) mass profile for the lens, which is expected to be a suitable model for galaxy scale lenses.

\subsubsection{Estimating time-delay from temporally unresolved images}
Consider a (micro)lensing scenario where the source is a stellar-mass CBC, and the lens is an intermediate mass black hole (IMBH). Here, since $\lambda \sim R_s$ the lensed GW images will not be temporally resolved, and wave-optics effects will distort the waveform with respect to the corresponding unlensed waveform \citep{lensing_gw_2, lensing_gw_4, microlensing, microlensing_cao_et_al, Dai_et_al_microlensing, microlensing_christian_et_al, microlensing_lai_et_al, microlensing_diego_et_al, Cheung_et_al_microlensing, Mishra_et_al_microlensing}.

GW templates incorporating the wave optics distortions (beating patterns) have been constructed, assuming the lens to be a point mass, as is adequate for the kind of microlensing scenario we consider here. Under the point mass approximation, the deformation of the signal depends on the redshifted mass of the lens ($M^z_l$), and the impact parameter ($y$). These can then be used to calculate the corresponding image time delay in the geometric optics approximation (which would have been observed if the images were temporally resolvable) \citep{microlensing}:

\begin{eqnarray}
    \Delta t_l = 4 M_{l}^z \left[ \frac{y \sqrt{y^2 + 4}}{2} + \ln \left( \frac{\sqrt{y^2 + 4} + y}{\sqrt{y^2 + 4} - y} \right)\right]
\end{eqnarray}

Indeed, if the CBC's GWs and FRB were both microlensed by the same IMBH, one distorted GW image and two temporally resolved FRB images would be observed. Comparing the time delay estimated from the distorted GW with that measured from the arrival times of the FRB images would then ascertain their common provenance.

\subsection{Association of FRBs with CBCs}
LIGO-Virgo-KAGRA (LVK) collaboration has searched for temporal and spatial coincidences of FRBs detected by CHIME (Canadian neutral Hydrogen Mapping Experiment) \citep{CHIME, CHIME_frb_catalog1} with CBCs detected via GWs during the third observing (O3) run \citep{LVK_CBC_FRB} of the LIGO-Virgo detectors. There was no confident association found for any of the FRBs with any of the CBCs. But the associations can not be ruled out either due to limitations in the precision on the measurement of luminosity distance (of CBCs and FRBs) and sky-location (of CBCs). 

The measurement of luminosity distance of FRBs is based on inferring the dispersion measure (DM) which is a measure of integrated column electron density along the line of sight \citep{distance_DM_frb, distance_uncertainties_FRB}. The conversion of DM to a luminosity distance will suffer from uncertainties and biases due to poor understanding of the matter-distribution in the intergalatic medium \citep{DM_Xu_Zhang, DM_uncertainties_Xu_et_al, LVK_CBC_FRB}. The luminosity distance and sky-location both are poorly measured in GW observations making the association even more difficult. For example, in the absence of any other messenger than GWs, with current detector sensitivities, the sky-localisation and luminosity-distance errors for compact binaries are typically $\sim \mathcal{O}(10 - 500 \rm{\ sq. deg.})$ \citep{Fairhurst_2009, Singer_2014, Singer_2016} and $\Delta d_L / d_L  \sim \mathcal{O}(1 - 10) $ \citep{distance_uncertainties_GW} respectively. Furthermore, the time of occurrence of the FRBs in the CBC's evolution is unknown, with some models predicting that FRBs occur as precursors, while others predict their occurrence postmerger. This necessitates a wide time window surrounding the CBC's merger time to search for temporal coincidences with FRBs, thus increasing the uncertainty in making a CBC-FRB association.

In this work, we discuss the possibility of searching for these associations using the gravitational lensing of FRBs and GWs from CBCs. If both FRB and GWs emitted by a compact binary merger are strongly lensed by a galaxy or galaxy cluster, it will lead to two or more identical copies of these signals separated by time-delays and signal amplitudes rescaled. If we can identify the lensed FRBs and GWs independently, we will be able to measure time-delays separately for the FRB-images and the GW-images. Instead of looking for coincidences of arrival times of the GWs and the FRB signal, matching these independently measured time-delays from FRB-images and GW-images can provide us with a more robust association. The probability that a lensed FRB is falsely associated with a lensed GW, at random, will be small enough to enable an unambiguous association or a confident dismissal. FRBs are also lensed by the plasma in between the source and the observer leading to additional time-delay between the different images of FRBs but these time-delays ($\sim 1 \mu\rm{s} - 10 \rm{ms}$) might not be significant enough to affect the strong lensing time-delay between the images \citep{frb_plasma_lensing1, frb_plasma_lensing2}.

\subsubsection{False association probability}
Let us assume that a lensed FRB/GW has a time-delay $\Delta t_l$ and the GW time-delay is measured with an uncertainly $\sigma_{\rm{GW}}$. The timing measurement in FRBs is relatively precise ${O}(\rm{ns})$, and hence we assume it to be a point estimate. We assume a positive association when time delays of lensed FRB and GW images agree when within the measurement uncertainty. The probability that a lensed FRB will have a time delay $\Delta t_l$ that is accidentally consistent with the time delay of a lensed GW is given by:
\begin{eqnarray}
    p = \int_{\Delta t_l - \sigma_{\rm{GW}}}^{\Delta t_l + \sigma_{\rm{GW}}} p(\Delta t_{\rm{GW}}) d\Delta t_{\rm{GW}}
\end{eqnarray}
where $p(\Delta t_{\rm{GW}})$ is the expected time-delay distribution for detected lensed GW events in a given observing scenario. If ther are $N$ lensed FRBs detected, the probability that at least one FRB will be falsely associated  (false association probability, FAP) 
 with one lensed GW is:
\begin{eqnarray}
    \mathrm{FAP} = 1 - ( 1 - p)^N,
\end{eqnarray}
When $p$ is very small, 
\begin{eqnarray}
    \mathrm{FAP} \approx N p.
    \label{eq:FAP_expanded}
\end{eqnarray}
It is obvious from Eq. (\ref{eq:FAP_expanded}) that FAP will increase linearly with detections of lensed FRBs.

\subsection{Rate of the associations}
Let $N_{\rm{tot}}$ be the total number of CBCs, and their redshift distribution probability be $dP_b / dz (z)$. Assuming the lensing probability to be $P_L (z)$, we can compute the number of lensed events as:
\begin{eqnarray}
    N_L = N_{\rm{tot}} \int_0^{z_{\rm{max}}} \frac{dP_b}{dz} (z) P_L (z) dz,
\end{eqnarray}
where $z_{\rm{max}}$ is the maximum redshift upto which the CBCs are distributed. Accounting for selection effects resulting from finite horizon distances of GW detectors, the number of detected lensed events is evaluated as:
\begin{eqnarray}
    N_{DL} = N_{\rm{tot}} \int_0^{z_{\rm{max}}} \frac{dP_b}{dz} (z) P_L (z) \bar{S}(z) dz,
\end{eqnarray}
where $\bar{S}(z)$ captures these selection effects as a function of redshift, averaged over rest of the binary parameters (e.g. component masses, spins, sky location, orientation, polarisation angle, etc.)~
Thanks to the (nearly) all-sky sensitivity of GW detectors and the non-interacting nature of gravitational radiation, it is straightforward to compute the selection effects of GW observations. We inject theoretical models of GW signals into Gaussian noise with appropriate power spectral densities of various detectors and identify signals crossing a network SNR of 8. The fraction of detected signals as a function of the source redshift $z$ provides a good approximation of the selection function $\bar{S}(z)$. 
If we assume that all CBCs produce FRBs and the horizon distance of GW and FRB detectors are approximately similar \citep{LVK_CBC_FRB}~\footnote{This is strictly not true for current sensitivities of LVK detectors. The radio detectors are sensitive to larger redshifts than GW detectors \citep{LVK_CBC_FRB}. But this assumption will hold, at least for a few upcoming generation of GW and radio detectors for the horizon distance ($z_{\rm{max}} = 6$) we have considered in this work.}, the lensing time-delay distribution for lensed and detected FRBs will be the same as that of strongly lensed and detected GWs. The number of detected lensed GWs/FRBs available for associations is then evaluated as:
\begin{eqnarray}
    N^{\rm{GW/FRB}}_{DL} = N_{\rm{tot}} \ f_{\rm{FOV}} \int_0^{z_{\rm{max}}} \frac{dP_b}{dz} (z) P_L (z) \bar{S}(z) dz,
\end{eqnarray}
where $f_{\rm{FOV}}$ is the fraction of detectable-lesned FRBs that will be seen by the radio telescopes due to limited field of view (FOV). FRB detectors (radio telescopes) will not be able to able to observe all time-delays due limited duration spent by the FRB-source in the FOV of the telescope. This will introduce another selection effect in the time-delay distribution observed for lensed FRBs (see Appendix \ref{app:frb_FOV_selection_effects} for details). The fraction of CBC-FRB associations to be detected:
\begin{eqnarray}
    f_{\rm{GW-FRB}} = \frac{N^{\rm{GW/FRB}}_{DL}}{N_{\rm{tot}}}.
\end{eqnarray}
Our goal is to predict the number of CBC-FRB associations to be detected per year. Let us assume the rate of mergers per unit redshift and per unit observer time is given by:
\begin{eqnarray}
    \frac{d^2N}{dz dt_{\rm{obs}}} (z) = \mathcal{R} (z) \frac{dV_c}{dz} \frac{1}{1+z}.
\end{eqnarray}
Here $\mathcal{R} (z)$ denotes the merger rate  as a function of redshift $z$ per unit comoving volume $V_c$ per unit source frame time. The merger rate $\mathcal{R} (z)$ can be decoupled into a merger rate at $z=0$ as $\mathcal{R}_0$ (local merger rate) and a part with $z$ dependence as $\psi(z)$. Then:
\begin{eqnarray}
    \mathcal{R} (z) = \mathcal{R}_0 \psi(z),
\end{eqnarray}
where $\psi(z)$ is chosen such that $\psi (0) = 1$ i.e. $\mathcal{R} (0) = \mathcal{R}_0$. We would like to compute the number of mergers occurring per unit time ($\tilde{\mathcal{R}}$) up to a maximum redshift $z_{\rm{max}}$ as:
\begin{eqnarray}
    \tilde{\mathcal{R}} = \mathcal{R}_0 \int_0^{z_{\rm{max}}} \psi(z) \frac{dV_c}{dz} \frac{1}{1+z} \ dz,
\end{eqnarray}
Then the number of CBC-FRB associations to be detected per year is given by
\begin{eqnarray}
    N_{\rm{Asso, det}}^{\rm{CBC-FRB}} = f_{\rm{GW-FRB}} \times \tilde{\mathcal{R}}.
\end{eqnarray}
Please Note that the number of CBC-FRB associations will depend on the fraction of CBCs emitting FRBs. For CBC mergers, there are multiple mechanisms that could lead to FRB-like emission. It is difficult to comment on what fraction of CBCs will produce FRBs given the strength of the magnetic field of NSs and charge of BHs in a binary are not very well understood. While we assume that all CBC mergers emit FRBs, our final estimates of the lensing association rates can simply be rescaled for any value of this fraction.

\subsection{GW detector networks and radio telescopes}
We consider here four observing scenarios for gravitational waves: (i) O4 --- 4 detector network [two LIGOs operating at design (aLIGO) sensitivity, Virgo, and KAGRA] with sensitivities corresponding to the fourth observing run \citep{AdvLIGO_Virgo, advligo, advvirgo}, (ii) O5 --- 5 detector network --- three LIGOs, including LIGO-India, at A+ sensitivities, Virgo and KAGRA at aLIGO sensitivities) \citep{kagra, observer_summary, SaleemLIGOIndia}, (iii) Voyager -- 5 detector network with all three LIGO detectors upgraded to Voyager sensitivities \citep{Adhikari_voy, Voyager_PSD, instrument_science}, and (iv) 3G ---- 3 detector network with one Einstein Telescope \citep{ET, ET_PSD} and two Cosmic Explorers \citep{CE_PSD, CE}. 

On the radio observation front, we consider the currently operational CHIME \citep{CHIME} and the proposed Bustling Universe Radio Survey Telescope (BURSTT) in Taiwan \citep{BURSTT}. The CHIME has a FOV of about 200 sq.deg. while BURSTT can observe sources from horizon to horizon, with a FOV of around 10,000 sq.deg. We also consider a scenario of an array of radio telescopes with sensitivities similar to SKA \citep{SKA} that can scan the whole sky at given time.

%% file: results.tex
We consider a population of 110 million non-spinning NSBH binaries and BNSs with a redshift distribution closely following the star formation history \citep{madau_dickinson}. The BH and NS mass distributions are assumed to be log-uniform and uniform respectively. The mass range spanned by BHs and NSs in our analysis are: $m_{\rm{BH}} \in [5, 100] M_{\odot}$ and $m_{\rm{NS}} \in [1, 3] M_{\odot}$. The binaries are distributed uniformly in the sky ($\alpha, \sin\delta$) and polarization sphere ($\cos\iota, \psi$). 

The gravitational lenses are drawn from the galaxy-population observed by Sloan Digital Sky Survey (SDSS) \citep{sdss, sdss_old}. We compute the optical depth (lensing probability) assuming the SIE mass density profile, which is expected to adequately model galaxies \citep{kormann_et_al}. Under this assumption images can either occur as doubles (two images) or quads (four images). The number of images depends on the location of the source in the source-plane with respect to the caustics -- curves in the source-plane where the magnification formally diverges.

\begin{figure}
    \centering
    \includegraphics[scale=0.48]{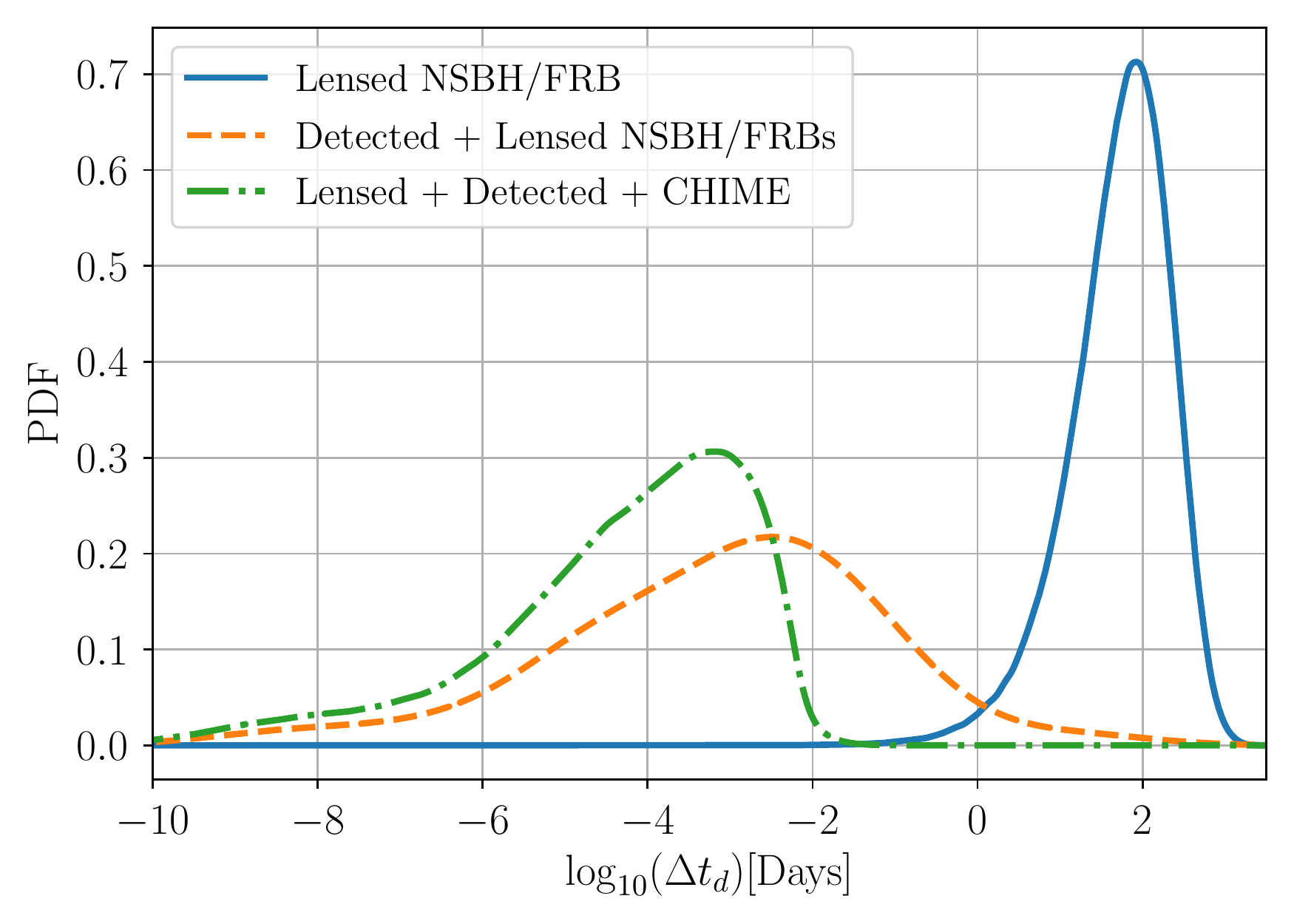}
    \caption{The figure shows the time delay distribution of all lensed NSBH mergers (blue solid line), and a subset detected by GW detectors (orange dashed line) in O4 observing scenario. The lensed FRBs will have the same time-delay distributions when the selection effects due to finite FOV of the radio telescope is omitted. The green dashed-dotted line corresponds to the subset of lensed mergers that are detectable by GW detectors as well as the CHIME radio telescope.}
    \label{fig:time_delay_distr_O4_CHIME}
\end{figure}

Assuming geometric optics, we infer the positions of the GW images ($x_i$) as well as magnifications ($\mu_i$) \citep{haris_et_al}. The lensed GW signal is considered as detected if two or more images have network signal-to-noise ratio (SNR) greater than 8. We compute the time delays for the different images with respect to the one arriving first at the detector. It is clear from Fig. \ref{fig:time_delay_distr_O4_CHIME} that larger time delays are selected out when applying the detection criteria since they correspond to the sources located at larger distances. The lensed-detected time-delay distribution is dominated by small time  delays that corresponds to the brightest image pairs which reside near the caustic. Most of these pairs correspond to the second and third images in quads that are typically brighter than the first and the fourth, and have small time delays (see, e.g., \cite{magare2023}).

\begin{figure}
    \centering
    \includegraphics[scale=0.43 ]{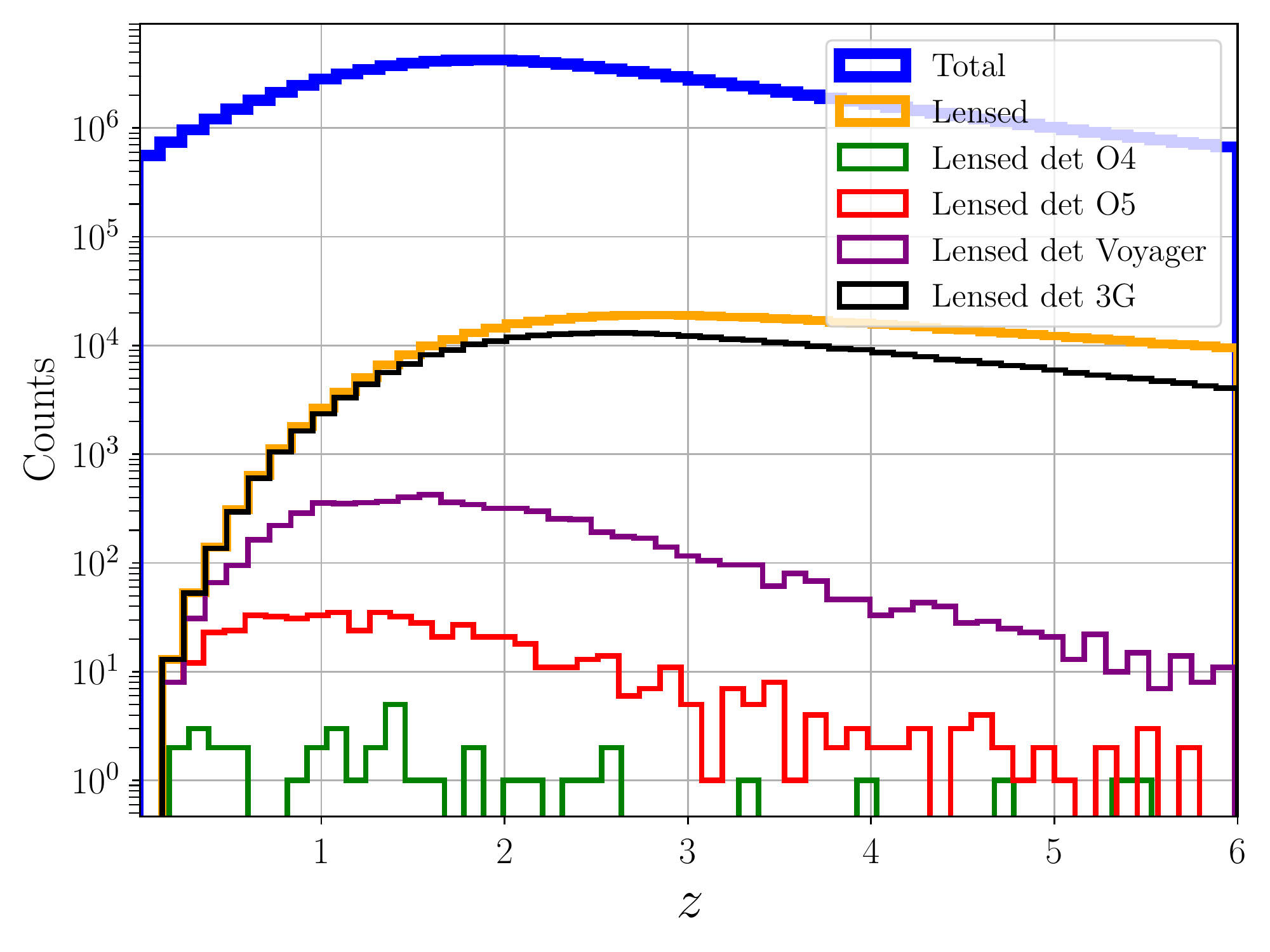}
    \caption{The distributions of total NSBH injections along with the lensed and detected-lensed ones in different observing scenarios. We do not consider any mergers with $z > z_{\rm{max}} = 6$. The effect of this $z_{\rm{max}}$ cutoff is negligible in O4, O5, and Voyager observing scenarios. However, in 3G some of the high redshift mergers ($z > 6$) could be magnified due to lensing, rendering them detectable. We have not considered this effect here. }
    \label{fig:hist_NSBH_lensed_source_redshifts}
\end{figure}

Assuming that both GW detectors and radio telescopes have similar horizon distance, the time-delay distribution will be similar for both GW and FRB events. However, since radio telescopes, such as the ones we focus on in this work, viz. CHIME \citep{CHIME} and BURSTT \citep{BURSTT}, are not all-sky like GW detectors but have a limited FOV, mostly FRBs with relatively smaller time-delays will be detectable. 

In O4, the detectable lensing fraction for NSBHs is low ($\sim 6 \times 10^{-5} \%$) since most of the lensed events lie beyond the horizon distance of GW detectors (see Fig. \ref{fig:hist_NSBH_lensed_source_redshifts}). Those that are detectable have lower luminosity distances, and thus produce smaller time delays (see Fig. \ref{fig:time_delay_distr_O4_CHIME}). On the other hand, not all lensed FRBs corresponding to detectable lensed GWs are observable in radio since the radio telescopes can only see a patch of the total sky at a time due to their moderate FOV. This further limits the distribution of detectable time delays to smaller values. 
Accounting for all these selection effects and assuming the NSBH merger rates (at $90\%$ confidence) estimated from O3a (first half of the third) observing run of LIGO-Virgo-KAGRA (LVK) detector network as $\sim 7.8 - 140 \ \rm{Gpc}^{-3} \rm{yr}^{-1}$, we find the rates of NSBH-FRB associtation to be detected in O4+CHIME era to be $\sim 5 \times 10^{-5} - 10^{-3} \ \rm{yr}^{-1}$~\footnote{The beaming of FRBs can further reduce the detectable fraction leading to lower association rates. But the beaming angle for the FRBs emitted from compact binary mergers is largely unconstrained \citep{frb_beaming}. So we neglect this in our analysis. The rates provided in this paper correspond to the upper limits for CBC-FRB associations.}. This suggests that it is unlikely that such an association can be made in the upcoming (fourth) observing run of the LVK.

\begin{figure*}
    \centering
    \includegraphics[scale=0.47]{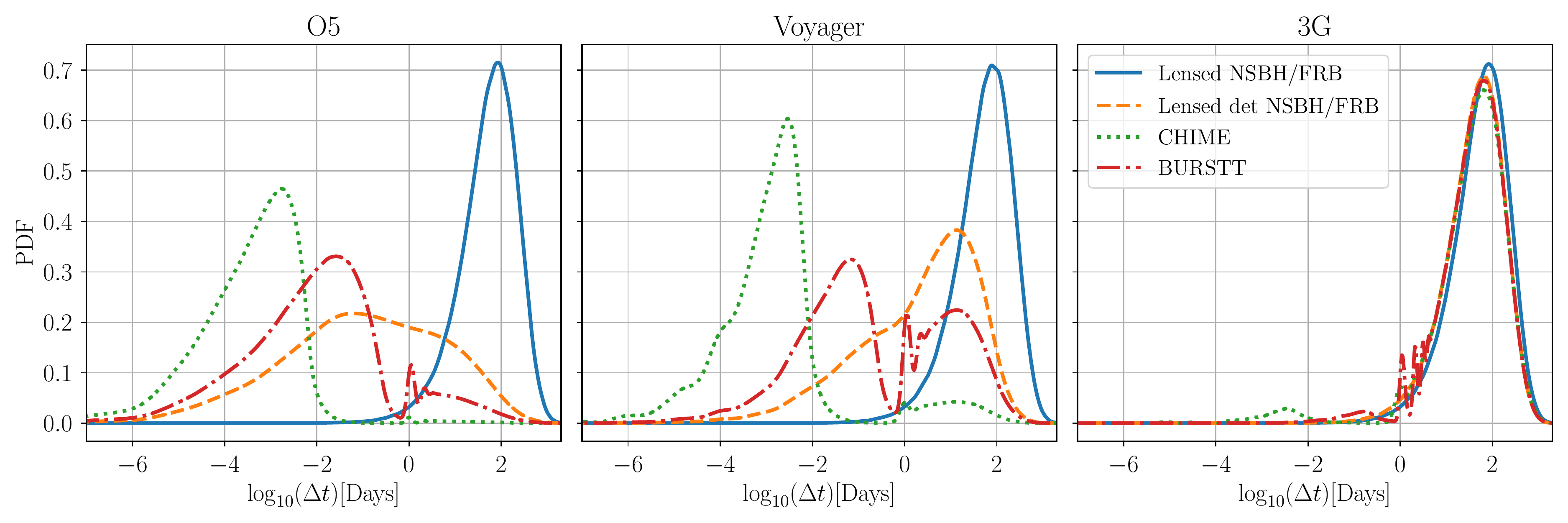}
    \caption{Time-delay distributions for lensed NSBH mergers as in Fig. \ref{fig:time_delay_distr_O4_CHIME} but for various future GW observing scenarios --- O5 (left), Voyager (middle), and 3G (right) --- with CHIME and BURSTT radio telescopes considered for observation of lensed FRBs.  It is obvious that BURSTT radio telescope, which has a larger FOV, can observe longer time-delays than CHIME and hence increases the fraction detected-lensed FRBs. The peak at one day in the time-delay distribution seen by the BURSTT telescope is due to the rotation period of radio telescope fixed on earth. There are further peaks at the integral multiple of days but due to log-scale, they appear as a continnum for time delays much greater than a day.}
    \label{fig:time_delay_O5_Voy_3G_CHIME_BURSTT}
\end{figure*}

\begin{figure*}
    \centering
    \includegraphics[scale=0.47]{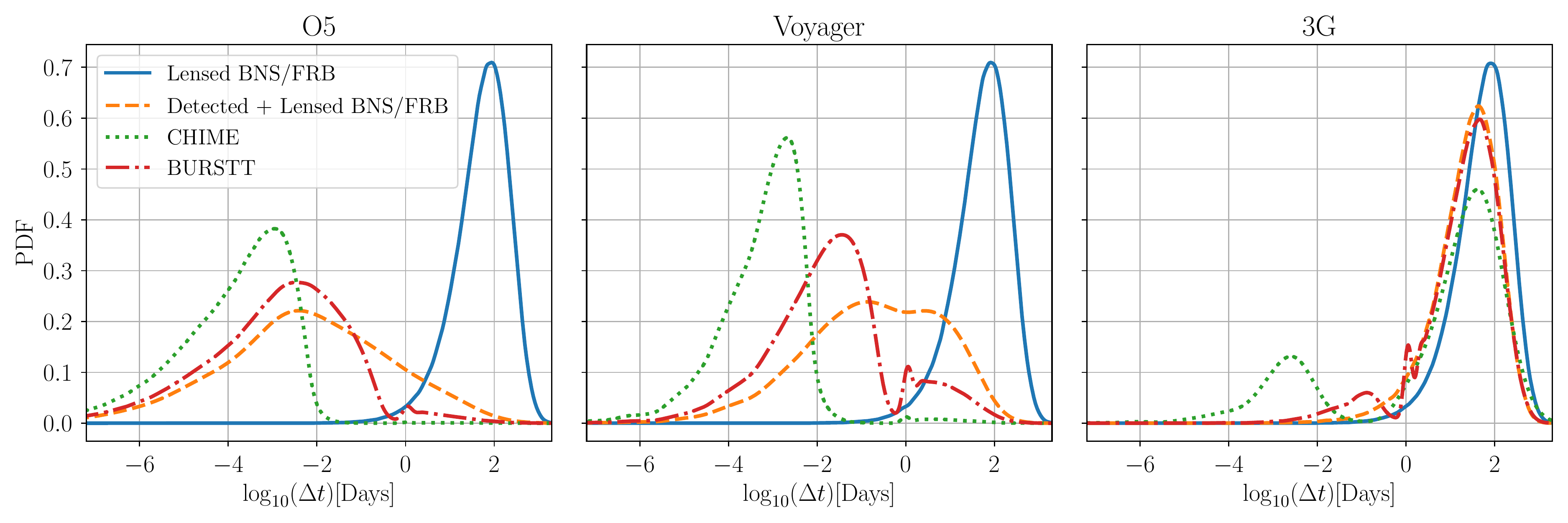}
    \caption{ Same as Fig. \ref{fig:time_delay_O5_Voy_3G_CHIME_BURSTT} but the time-delay distributions correspond to BNS mergers.}
    \label{fig:time_delay_BNS_O5_Voy_3G_CHIME_BURSTT}
\end{figure*}

Considering future observing scenarios of GW detectors with improved sensitivities, we find that the detected lensing fraction increases due to observability of sources at high redshifts. Consequently, the time-delay distribution for detected-lensed GWs/FRBs shifts towards longer time-delays. We find the detectable lensing fraction of NSBH mergers to be $\sim 5 \times 10^{-4} (6 \times 10^{-3}) [0.3] \%$ in O5 (Voyager) [3G] observing scenario (Fig. \ref{fig:hist_NSBH_lensed_source_redshifts}). Focusing on CHIME radio telescope \citep{CHIME}, 
the rate of such NSBH-FRB association is estimated to be $\sim 2 \times 10^{-4} - 3 \times 10^{-3} (4 \times 10^{-4} - 10^{-2}) [3 \times 10^{-3} - 5 \times 10^{-2}] \ \rm{yr}^{-1}$ in O5 (Voyager) [3G]. Here, we account for the fact that CHIME beam will return to its original position in the sky after 24 hours. So the time delays which are greater than 24 hours can be detectable by CHIME if the time-delay modulo 24 hours is smaller than the source transit time in the FOV of the telescope. The contribution due to this time delay wrapping factor will be significant for more sensitive detectors (which will observe lensed CBC/FRB events out to large redshifts, which in turn will have larger time delays). This can be observed by the strength of the second broad peak in the time-delays seen by a radio telecope, e.g. see Fig. \ref{fig:time_delay_O5_Voy_3G_CHIME_BURSTT}. Similarily, with the consideration of BURSTT radio telescope \citep{BURSTT}, we find that the fraction of detected lensed FRBs increases. This corresponds to the rate of NSBH-FRB associations to be $\sim 0.02 - 0.5 (0.2 - 3) [5 - 87] \rm{yr}^{-1}$. The significant increase in the association rates is attributed to wrapping of time-delays by 24 hours which is prominent in more sensitive GW detector scenarios as well as the larger FOV of BURSTT telescope~\footnote{Apart from the two broad modes of time-delay distribution detected by radio telescopes, a few sharp peaks near time-delays of $\sim 1, 2, 3, ...$ days are visible in Fig. \ref{fig:time_delay_O5_Voy_3G_CHIME_BURSTT} and \ref{fig:time_delay_BNS_O5_Voy_3G_CHIME_BURSTT}. These are due to the fact that the probability of seeing smaller time-delays by a radio telescope (in case of BURSTT $\sim \mathcal{O}(10)$ minutes) is larger near the integral multiple of a day. In fact, the second broad peak is a collection of peaks at integral multiple ($>2$) of days condensed due to log scale.}.  When we consider BNS instead of NSBH mergers as the source of FRBs, the rate of BNS-FRB associations turns out to be $\sim 10^{-4} - 2 \times 10^{-2} (4\times 10^{-4} - 7 \times 10^{-2})[2 \times 10^{-3} - 0.3] \rm{yr}^{-1}$ and $\sim 9 \times 10^{-3} - 2 (7 \times 10^{-2} - 12)[3 - 422] \rm{yr}^{-1}$ in CHIME and BURSTT observing scenario respectively. We have assumed a BNS merger rate $\sim 10 - 1700 \ \rm{Gpc}^{-3} \rm{yr}^{-1}$ based on LVK GW observations in O3. Since the finite FOV of radio telescopes (e.g. CHIME) limits the probability of observing an FRB-CBC association. This motivates the need of an array of BURSTT like radio telescopes that can scan the whole sky in one time. The rate of associations in such case can be as large as $\sim \mathcal{O}(30)\ \left[\mathcal{O}(1800)\right] \ \rm{yr}^{-1}$ for NSBHs and $\sim \mathcal{O}(100)\ \left[\mathcal{O}(8500)\right] \ \rm{yr}^{-1}$ for BNSs in Voyager[3G] scenario. 

\begin{center}
\begin{table*}
\caption{The BNS/NSBH-FRB association rates has been tabulated in different GW (columns) and radio (rows) observing scenarios.}
\begin{tabular}{|c|ccc|}
\hline
\multirow{2}{*}{} & \multicolumn{3}{c|}{BNS (NSBH) {[} in $\rm{yr}^{-1}${]}}  \\ \cline{2-4} 
                  & \multicolumn{1}{c|}{O5} & \multicolumn{1}{c|}{Voyager} & 3G \\ \hline
CHIME &
  \multicolumn{1}{c|}{$10^{-4} - 2 \times 10^{-2} (2 \times 10^{-4} - 3 \times 10^{-3})$} &
  \multicolumn{1}{c|}{$4 \times 10^{-4} - 7 \times 10^{-2} (4 \times 10^{-4} - 8 \times 10^{-3})$} &
  $2 \times 10^{-3} - 0.3 (3 \times 10^{-3} - 5 \times 10^{-2})$ \\ \hline
BURSTT &
  \multicolumn{1}{c|}{$7 \times 10^{-3} - 2 (0.02 - 0.5)$} &
  \multicolumn{1}{c|}{$7 \times 10^{-2} - 12 (0.2 - 3)$} &
  $3 - 422 (5 - 87)$ \\ \hline
All Sky &
  \multicolumn{1}{c|}{$0.04 - 8 (0.2 - 3)$} &
  \multicolumn{1}{c|}{$0.6 - 100 (2 - 35)$} &
  $50 - 8580 (101 - 1814)$ \\ \hline
\end{tabular}

\label{tab:association_rates}
\end{table*}
\end{center}

\begin{figure}
    \centering
    \includegraphics[scale=0.42]{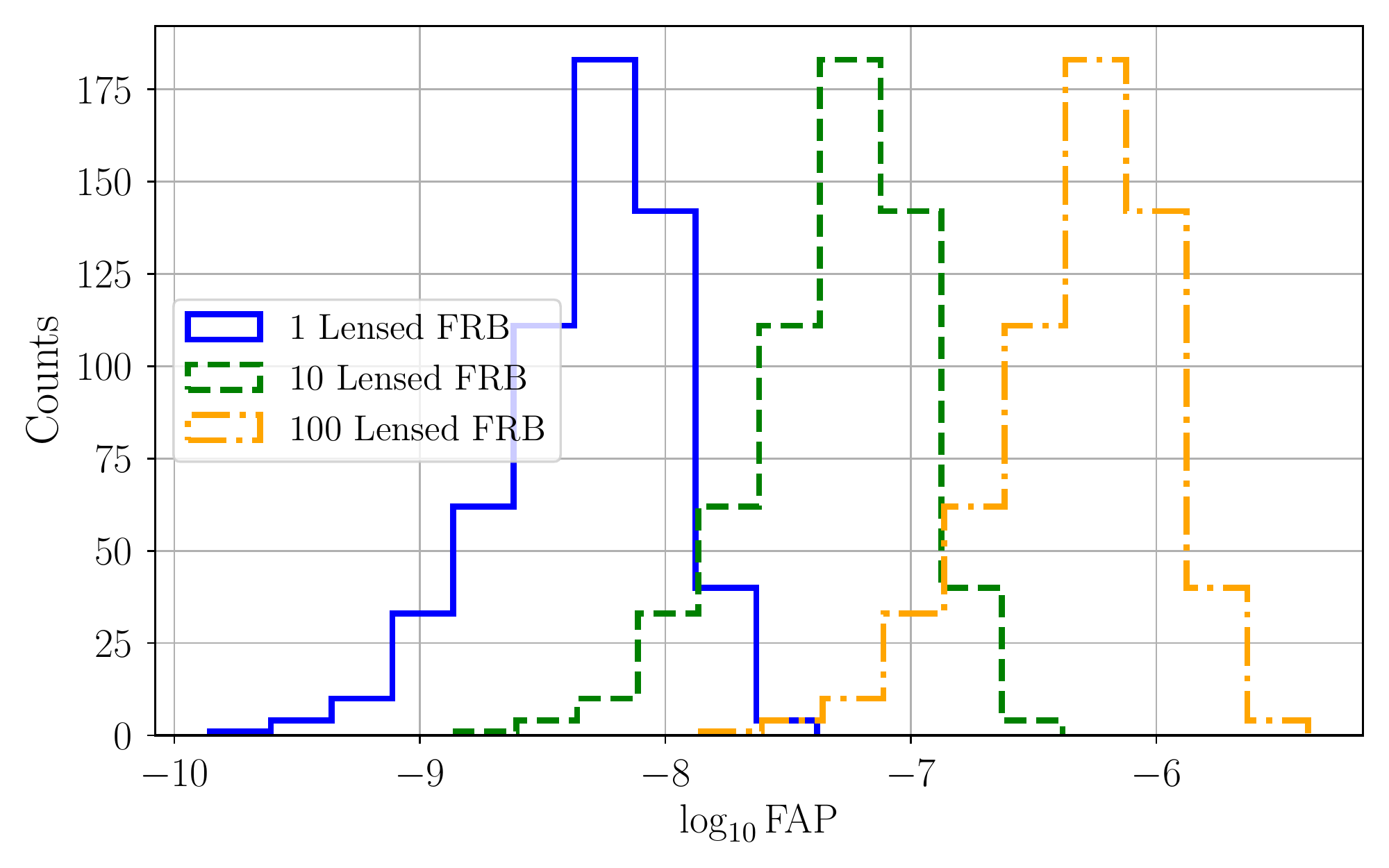}
    \caption{The figure shows the distribution of False association probabilities (FAPs) of lensed FRBs to be associated randomly with GW signals due to uncertainty in the measurement of lensing time-delays in gravitational-waves in O5. The FAP scales linearly with the number of lensed FRBs.}
    \label{fig:FAP_O5}
\end{figure}

We compute the FAP of randomly associating a detected lensed FRB with lensed detected GWs pairs. The CBC-FRB association can be made at $> 5\sigma (\rm{FAP} \sim 10^{-8})$ confidence given that one lensed FRB has been detected (see Fig. \ref{fig:FAP_O5}). The association confidence is essentially independent of the observing scenario since it depends on the timing uncertainties for GW signals which does not change a lot. However, the FAP will increase when several lensed FRBs are detected in a given observing scenario. This will reduce the confidence with which they can be associated with CBCs. Nevertheless, the association will be confident enough to verify or rule out the possibility of compact binary mergers being the source of FRBs.

\begin{figure}
    \centering
    \includegraphics[scale=0.33]{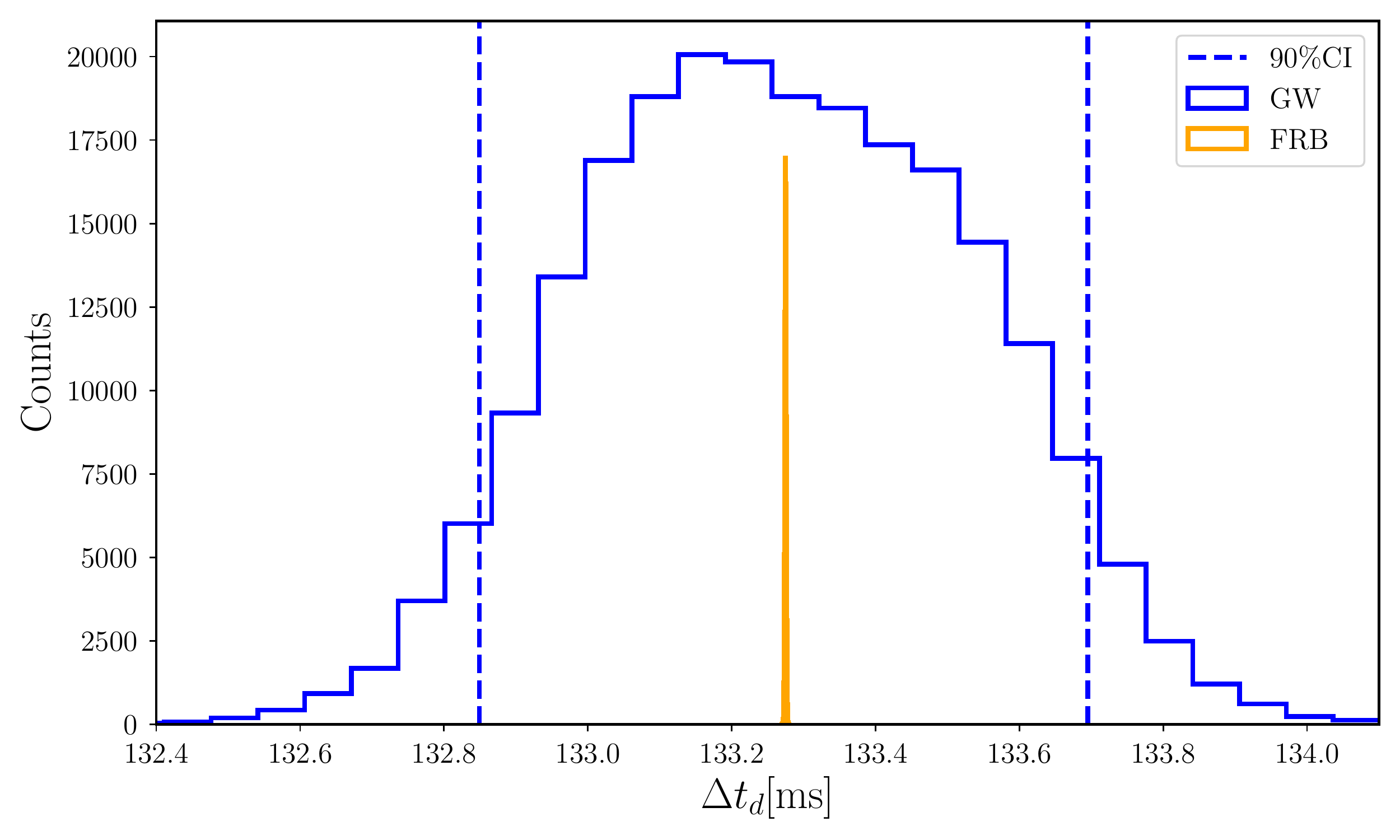}
    \caption{The figure shows an example of the time-delay posteriors inferred for a microlensed NSBH (blue) and corresponding lensed FRB (orange) is emitted. The time measurement for FRBs is quite accurate hence quite narrow posterior. It is clear that if one sees a microlensed GWs and corresponding strongly lensed FRB, matching the time-delays measured independently can be used to find out an association between them.}
    \label{fig:nsbh_microlensing_posterior}
\end{figure}

We also give an illustrative example which leverages the microlensing of an NSBH by an IMBH, modelled as a point mass lens. The distortions caused by the microlensing on the GW waveform \citep{microlensing} can be used to infer the lensing time-delay in the geometric optics limit. \citep{lensing_otto, lensing_LVK_O3a}. The Bayesian parameter estimation posterior for the time-delay inferred from an example NSBH merger has been shown in Fig. \ref{fig:nsbh_microlensing_posterior} along with 90\% credible intervals. If this NSBH merger emitted an FRB~\footnote{There are models suggesting that BBH mergers can also emit FRBs-like signal if at least one of the BHs have charge greater than some threshold value to produce the large luminosities observed for FRBs \citep{frb_charged_BH}.}, it will be lensed and have two temporally resolved images whose time-delay can be measured with a precision of $O(\rm{ns})$. We refrain from estimating the association rates of such events because it is strongly dependent on the number density, mass spectrum and redshift distribution of IMBHs (or, in general, MACHOs whose gravitational radii are of the order of the GW wavelength of stellar mass CBCs) lenses which to date are not well constrained.

%% file: conclusion.tex
The origin of FRBs is still an open question even after 15 years since the first discovery in 2007 by Parkes radio telescope in Australia \citep{lorimer_FRB}. Since then we have observed a plethora of FRBs with the help of various radio telescopes across the globe including the CHIME telescope \citep{frb_catalog_thornton_et_al, frb_arecibo, frb_masui_et_al, frbs_champion_et_al, frb_catalog_petroff_et_al, frb_caleb_et_al, frb_bannister_et_al, frb_shannon_et_al, CHIME_frb_catalog_old,CHIME_frb_catalog1}. Some of the FRBs are found repeating without any clear periodicity \citep{repeating_frb1, repeating_frb2, repeating_frb3}. A recent association of FRBs with a highly magnetized rotating young neutron star \citep{frb_magnetar1, frb_magnetar2} suggests that such magnetars could produce FRBs. But magnetars can not solely explain the relatively large volumetric rate of FRBs as well as surprisingly large luminosities for most of the FRBs \citep{frb_magnetar1}. 

CBCs are also speculated to be possible sources of such radio bursts \citep{frb_BNS_circumbinary_plasma, frb_precursor_BNS, frb_BNS_merger, frb_supramassive_NS1, frb_supramassive_NS2, frb_BNS_inspiral, frb_charged_BH, frb_charged_BHNS, frb_review_zhang, frb_BNS_zhang}. But the association of FRBs with CBCs is limited by two factors, among others \citep{LVK_CBC_FRB}. One of them is the poor measurement of the source-luminosity distance which is uncertain in both GWs measurements from CBCs \citep{distance_uncertainties_GW} and DM measurements in FRBs \citep{DM_Xu_Zhang, DM_uncertainties_Xu_et_al, LVK_CBC_FRB}. Second is the sky-localization of the source. While FRBs have sky localisation errors of ($\mathcal{O}(1)$ sq. deg.), GW localisation inference produces errors of ($\mathcal{O}(100)$ sq. deg.) with current detectors \citep{Fairhurst_2009, Singer_2014, Singer_2016}. Furthermore, the time of emission of the FRB during the CBC's evolution is also uncertain \citep{NS_collapse_times}. This makes the temporal association of CBCs with FRBs difficult. 

In this work, we propose an astrophysical scenario where searching for associations of FRBs and GWs emitted by CBCs could be drastically enhanced if both the signals are gravitationally lensed~\footnote{The problem of identifying lensed GWs and FRBs has been explored extensively. For strong lensing, the optical depth $\tau_{\rm{s}}(z)$ is rather small if the sources are located at moderate distances \citep{multiple_cosmo_lensing, cosmo_lensing2}. This reduces the chances of seeing multiple images of lensed GW/FRBs significantly with the current generation detectors. On the GW front, lensing searches have found no evidence for lensing in the GW signals observed by LIGO-Virgo detectors \citep{lensing_otto, Alvin_et_al_lensing_search_LV, Dai_et_al_lensing_search_O2, Connor_et_al_lensing_search_O1O2, Pang_et_al_lensing_search_GW, Liu_et_al_lensing_search_O2, lensing_LVK_O3a}. On the FRB front, the lack of distinguishable morphological features in lensed FRB images limits the lensing search, but some clever techniques, like searching for phase-coherence at the electric field level, have been developed recently to identify them \citep{microlensed_frbs_zarif, Leung_et_al_frb_lensing}. As of now, there have been no confident detection of lensed FRBs, assuming lens masses $\sim 10^{-4} - 10^4 M_{\odot}$ \citep{microlensed_frbs_zarif}.}. If we can detect two or more lensed images of both FRB and GW, it will allow us to measure the time-delays between the images separately for both the signals. We find that an association with $>5\sigma$ confidence can be made by matching these independently estimated image time-delays which are expected to be measured precisely for FRBs and GWs from CBCs. 

Assuming the rate estimates of NSBH mergers from O3b LVK run, and that all NSBHs produce FRBs, we find that FRB-NSBH associations can be made with a maximum rate of $\sim 10^{-3} \mathrm{yr}^{-1}$ in the O4+CHIME observing scenario. This highly suppressed rate is due to the following two reasons: (i) fraction of detected-lensed events is not very large, especially for NSBH and BNS mergers (ii) the radio telescopes have moderate FOV which limits the possibility of observing the longer time-delays for lensed FRBs. These selection effects will reduce when more sensitive GW detectors (e.g. O5, Voyager, 3G, etc.) and more sensitive and wide field radio telescopes (e.g. BURSTT, etc.) become operational in the future. We find that the rate of detected lensed NSBH-FRB associations is $\sim 0.02 - 0.5 (0.2 - 3) [5 - 87] \rm{yr}^{-1}$ in O5 (Voyager) [3G] observing scenario with BURSTT radio telescope in operation. Considering BNS mergers as the sources of FRBs, the rate of associations becomes $\sim 9 \times 10^{-3} - 2 (7 \times 10^{-2} - 12)[3 - 422] \rm{yr}^{-1}$. If there existed an array of BURSTT like radio telescopes but with sensitivies similar to SKA that can see the whole sky at a given time, the rates of such associations can be as high as $\sim \mathcal{O}(30)\ \left[\mathcal{O}(1800)\right] \ \rm{yr}^{-1}$ for NSBHs and $\sim \mathcal{O}(100)\ \left[\mathcal{O}(8500)\right] \ \rm{yr}^{-1}$ for BNSs in Voyager[3G] scenario. This will allow us to unambiguously make an association or to rule out CBC as sources of FRBs. We also demonstrate that a microlensed NSBH merger can be used to find association between GW signal with FRB but the rate of such associations will be very uncertain, due to the poor understanding of the abundance and mass distribution of IMBHs/MACHOs. 

Since we will detect a much larger number of unlensed CBCs/FRBs than lensed ones (lensing optical depth $\tau_s \sim 0.1 - 1\%$), one might wonder if it will be possible to establish a statistical association between a population of CBCs and FRBs, without relying on lensing. We think that lensing will have the best ability to establish/rule out the association despite being rare. As we have elaborated in this paper, $\emph{one}$ joint detection of a CBC-FRB lensed pair is sufficient to establish the CBC origin of the FRB. This is expected with Voyager GW detector operating along with BURSTT telescope. Indeed, we expect the detection of $\sim 1000$ unlensed GW-FRB signals by this time. One could potentially compare the angular correlation function of the GW events in the sky with the angular correlation function of FRBs to check their association. However, it is very unlikely that we will be able to establish an association using these $\sim 1000$ events, primarily due to the poor GW localisation. For example, past studies \citep{Aditya_V_large_scale} have shown that, in order to establish the statistical association between galaxies and BBH mergers, we will need to observe tens of thousands of BBHs with 3G detectors. The requirements for GW-FRB association are likely to be similar, however more detailed studies are needed to compare the relative merits of the two methods.

There are possible ways of improving the association of lensed FRBs and GWs. Checking for relative magnification consistency between lensed FRBs and GWs would improve the confidence of CBC-FRB association, but not significantly (primarily due to the poor measurement of the luminosity distance and hence magnification ratio from GW observations.)
Apart from searching for the CBC-FRB associations between the detected images of lensed GWs/FRBs, a subthreshold search can be carried out to look for counterpart lensed images. If a lensed GW is detected, and one super-threshold FRB is found to be within the coincident time-window of a GW, then a subthreshold search for the second FRB image (whose expected arrival time is known a-priori from the GW image time delay). Similarly, the detection of a lensed FRB and one super-threshold GW image can be leveraged to identify the second subthreshold GW image. Such searches would contribute to an increase in the CBC-FRB association rate.


In addition to enabling a CBC-FRB association, lensing could also shed light on the physics of the FRB emission from a CBC. In particular, it would indicate the exact time during the CBC's evolution when the FRB is emitted, which would help determine and constrain the mechanisms that produce FRBs. If multiple FRBs are produced during the CBC's evolution, corresponding models would be validated as well, thus potentially shedding light on the origin of a fraction of the FRB repeaters.

%% file: appendix.tex
\begin{appendix}
\section{Computing FRB selection effects for radio telescopes}
\label{app:frb_FOV_selection_effects}
\begin{figure}
    \centering
    \includegraphics[scale=0.4]{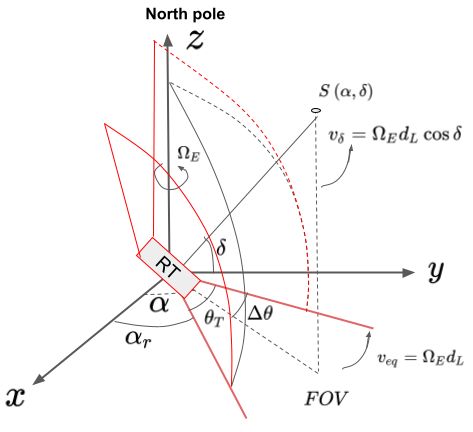}
    \caption{The figure shows the schematic to denote the selection effects due to the finite FOV of CHIME. The radio telescope, `RT', rectangular in shape have a FOV along the rectangular arc (red arc area) in the sky. It is clear that the rectangular arc will form a larger angle near the pole than at the equator. As it can be seen, the lensing time-delays detected by the radio telescope will depend on the location ($\alpha, \delta$) of the source.}
    \label{fig:CHIME_FOV}
\end{figure}

Considering the case of CHIME radio telescope, which is a collection of cylindrical meshes making a rectangular shape (denoted by RT in Fig. \ref{fig:CHIME_FOV}), the sources will appear moving East to West in its field-of-view (FOV). 
Let us denote the East to West FOV of the telescope by the angle $\theta_T$. Assume that at a given time, the source has a sky-location as $(\alpha, \delta)$. If we further assume that telescope is located at an angle $\alpha_r$ from the $x-$axis, the angular distance covered by the source in the FOV of the telescope
\begin{eqnarray}
    \Delta \theta = \theta_T - (\alpha - \alpha_r)
\end{eqnarray}
The distance travelled by the source with luminosity distance $d_L$
\begin{eqnarray}
    \Delta x = \Delta \theta d_L = \{ \theta_T - (\alpha - \alpha_r) \} d_L
\end{eqnarray}
The linear speed of the source with respect to earth
\begin{eqnarray}
    v_s = \Omega_E d_L \cos \delta
\end{eqnarray}
where $\Omega_E$ is angular speed of the Earth. The maximum time $\Delta t_{\rm{max}}$, the source will spend in the FOV of the telescope
\begin{eqnarray}
    \Delta t_{\rm{max}} = \frac{\Delta x}{v_s} = \frac{\theta_T - (\alpha - \alpha_r)}{\Omega_E \cos \delta}
\end{eqnarray}
The angle $\alpha_r$ is a reference for the telescope, we can choose it to be zero, then
\begin{eqnarray}
    \Delta t_{\rm{max}} = \frac{\theta_T - \alpha }{\Omega_E \cos \delta}
\end{eqnarray}
The only sources with $0 \leq \alpha \leq \theta_T$ will be seen by the telescope. It should be evident from the above expression that time spent by the source, with fixed right ascension ($\alpha$) will be different depending on the declination ($\delta$) of the source. This is because the linear speed of the source will be smaller than at the equator, or in other words, the angular distance covered by the source will larger near the poles than at the equator. Given an uniform distribution of FRBs in the sky, we will get a distribution of $\Delta t_{\rm{max}}$, i.e. $p(\Delta t_{\rm{max}}| \alpha, \delta)$. Using this distribution, we can compare it with the time-delay distribution ($p(\Delta t | \alpha, \delta)$) for detected lensed FRBs. We can statistically sample the time delays from both distributions and select the ones which satisfy the condition $\Delta t \leq \Delta t_{\rm{max}}$. Let us draw $N_{\rm{s}}$ samples from both $p(\Delta t_{\rm{max}}| \alpha, \delta)$ and $p(\Delta t | \alpha, \delta)$ randomly such that 
$$\Delta t^j \sim p(\Delta t | \alpha, \delta)$$, 
$$\Delta t_{\rm{max}}^j \sim p(\Delta t_{\rm{max}}| \alpha, \delta)$$ 
where $j \in \{1, 2, ..., N_{\rm{s}}\}$. Here, the symbol $\sim$ means ``drawn from''. Now collect the $\Delta t^j$ which satisfy the condition 
$$\Delta t^j \leq \Delta t_{\rm{max}}^j$$
Let us assume that the number of such $\Delta t^j$ is $N_{\rm{s}}^{\rm{det}}$. Then, the fraction of detected lensed FRBs
\begin{eqnarray}
    f_{\rm{FOV}} = \frac{N_{\rm{s}}^{\rm{det}}}{N_{\rm{s}}} \times f_{p}
\end{eqnarray}
where $f_p$ is the fraction of the total sky seen by the radio telescope at a given time 
\begin{eqnarray}
    f_p = \frac{\rm{FOV \ of \ CHIME}}{\rm{Total \ sky \ area}}
\end{eqnarray}
For CHIME \citep{CHIME}
\begin{eqnarray}
    f_p = \frac{200 \ \rm{sq. deg.}}{41235 \ \rm{sq. deg.}} = 0.00485
\end{eqnarray}
and for BURSTT \citep{BURSTT}
\begin{eqnarray}
    f_p = \frac{10000 \ \rm{sq. deg.}}{41235 \ \rm{sq. deg.}} = 0.2425
\end{eqnarray}

\end{appendix}

%% file: cbc_frb_driver.bbl
\begin{thebibliography}{}
\makeatletter
\relax
\def\mn@urlcharsother{\let\do\@makeother \do\$\do\&\do\#\do\^\do\_\do\%\do\~}
\def\mn@doi{\begingroup\mn@urlcharsother \@ifnextchar [ {\mn@doi@}
  {\mn@doi@[]}}
\def\mn@doi@[#1]#2{\def\@tempa{#1}\ifx\@tempa\@empty \href
  {http://dx.doi.org/#2} {doi:#2}\else \href {http://dx.doi.org/#2} {#1}\fi
  \endgroup}
\def\mn@eprint#1#2{\mn@eprint@#1:#2::\@nil}
\def\mn@eprint@arXiv#1{\href {http://arxiv.org/abs/#1} {{\tt arXiv:#1}}}
\def\mn@eprint@dblp#1{\href {http://dblp.uni-trier.de/rec/bibtex/#1.xml}
  {dblp:#1}}
\def\mn@eprint@#1:#2:#3:#4\@nil{\def\@tempa {#1}\def\@tempb {#2}\def\@tempc
  {#3}\ifx \@tempc \@empty \let \@tempc \@tempb \let \@tempb \@tempa \fi \ifx
  \@tempb \@empty \def\@tempb {arXiv}\fi \@ifundefined
  {mn@eprint@\@tempb}{\@tempb:\@tempc}{\expandafter \expandafter \csname
  mn@eprint@\@tempb\endcsname \expandafter{\@tempc}}}

\bibitem[\protect\citeauthoryear{Aasi et~al.}{Aasi et~al.}{2015}]{advligo}
Aasi J.,  et~al., 2015, \mn@doi [Classical and Quantum Gravity]
  {10.1088/0264-9381/32/7/074001}, 32, 074001

\bibitem[\protect\citeauthoryear{Abbott et~al.}{Abbott et~al.}{2017a}]{CE_PSD}
Abbott B.~P.,  et~al., 2017a, \mn@doi [Classical and Quantum Gravity]
  {10.1088/1361-6382/aa51f4}, 34, 044001

\bibitem[\protect\citeauthoryear{Abbott, Abbott, Abbott  et~al.}{Abbott
  et~al.}{2017b}]{gw170817}
Abbott B.~P.,  Abbott R.,  Abbott T.~D.,   et~al., 2017b, \mn@doi [Phys. Rev.
  Lett.] {10.1103/PhysRevLett.119.161101}, 119, 161101

\bibitem[\protect\citeauthoryear{{Abbott} et~al.}{{Abbott}
  et~al.}{2019}]{gwtc-1}
{Abbott} B.~P.,  et~al., 2019, \mn@doi [Physical Review X]
  {10.1103/PhysRevX.9.031040}, \href
  {https://ui.adsabs.harvard.edu/abs/2019PhRvX...9c1040A} {9, 031040}

\bibitem[\protect\citeauthoryear{{Abbott} et~al.}{{Abbott}
  et~al.}{2020a}]{gwtc-2}
{Abbott} R.,  et~al., 2020a, arXiv e-prints, \href
  {https://ui.adsabs.harvard.edu/abs/2020arXiv201014527A} {p. arXiv:2010.14527}

\bibitem[\protect\citeauthoryear{{Abbott} et~al.,}{{Abbott}
  et~al.}{2020b}]{BNS2_LIGO}
{Abbott} B.~P.,  et~al., 2020b, \mn@doi [\apjl] {10.3847/2041-8213/ab75f5},
  \href {https://ui.adsabs.harvard.edu/abs/2020ApJ...892L...3A} {892, L3}

\bibitem[\protect\citeauthoryear{{Abbott}, {Abbott}, {Abbott}  et~al.}{{Abbott}
  et~al.}{2020c}]{gw190425}
{Abbott} B.~P.,  {Abbott} R.,  {Abbott} T.~D.,   et~al., 2020c, \mn@doi [\apjl]
  {10.3847/2041-8213/ab75f510.48550/arXiv.2001.01761}, \href
  {https://ui.adsabs.harvard.edu/abs/2020ApJ...892L...3A} {892, L3}

\bibitem[\protect\citeauthoryear{{Abbott} et~al.}{{Abbott}
  et~al.}{2021a}]{gwtc-3}
{Abbott} R.,  et~al., 2021a, arXiv e-prints, \href
  {https://ui.adsabs.harvard.edu/abs/2021arXiv211103606T} {p. arXiv:2111.03606}

\bibitem[\protect\citeauthoryear{{Abbott}, {Abbott}, others, {Ligo Scientific
  Collaboration}, {VIRGO Collaboration}  \& {KAGRA Collaboration}}{{Abbott}
  et~al.}{2021b}]{LVK_NSBHs}
{Abbott} R.,  {Abbott} T.~D.,  others {Ligo Scientific Collaboration} {VIRGO
  Collaboration}  {KAGRA Collaboration} 2021b, \mn@doi [\apjl]
  {10.3847/2041-8213/ac082e10.48550/arXiv.2106.15163}, \href
  {https://ui.adsabs.harvard.edu/abs/2021ApJ...915L...5A} {915, L5}

\bibitem[\protect\citeauthoryear{{Abbott}, {LIGO Scientific Collaboration}  \&
  {Virgo Collaboration}}{{Abbott} et~al.}{2021c}]{lensing_LVK_O3a}
{Abbott} R.,  {LIGO Scientific Collaboration}  {Virgo Collaboration} 2021c,
  \mn@doi [\apj] {10.3847/1538-4357/ac23db}, \href
  {https://ui.adsabs.harvard.edu/abs/2021ApJ...923...14A} {923, 14}

\bibitem[\protect\citeauthoryear{Acernese et~al.}{Acernese
  et~al.}{2015}]{advvirgo}
Acernese F.,  et~al., 2015, \mn@doi [Classical and Quantum Gravity]
  {10.1088/0264-9381/32/2/024001}, 32, 024001

\bibitem[\protect\citeauthoryear{Adhikari et~al.}{Adhikari
  et~al.}{2019}]{Adhikari_voy}
Adhikari R.~X.,  et~al., 2019, \mn@doi [Classical and Quantum Gravity]
  {10.1088/1361-6382/ab3cff}, 36, 245010

\bibitem[\protect\citeauthoryear{Amiri et~al.,}{Amiri
  et~al.}{2018}]{amiri2018chime}
Amiri M.,  et~al., 2018, The Astrophysical Journal, 863, 48

\bibitem[\protect\citeauthoryear{Aso et~al.}{Aso et~al.}{2013}]{kagra}
Aso Y.,  et~al., 2013, \mn@doi [Phys. Rev. D] {10.1103/PhysRevD.88.043007}, 88,
  043007

\bibitem[\protect\citeauthoryear{{Bannister} et~al.,}{{Bannister}
  et~al.}{2017}]{frb_bannister_et_al}
{Bannister} K.~W.,  et~al., 2017, \mn@doi [\apjl] {10.3847/2041-8213/aa71ff},
  \href {https://ui.adsabs.harvard.edu/abs/2017ApJ...841L..12B} {841, L12}

\bibitem[\protect\citeauthoryear{{Bartelmann}}{{Bartelmann}}{2010}]{lensing_review}
{Bartelmann} M.,  2010, \mn@doi [Classical and Quantum Gravity]
  {10.1088/0264-9381/27/23/233001}, \href
  {https://ui.adsabs.harvard.edu/abs/2010CQGra..27w3001B} {27, 233001}

\bibitem[\protect\citeauthoryear{{Bhardwaj}, {Palmese}, {Maga{\~n}a Hernandez},
  {D'Emilio}  \& {Morisaki}}{{Bhardwaj}
  et~al.}{2023}]{off_axis_BNS_on_axis_FRB}
{Bhardwaj} M.,  {Palmese} A.,  {Maga{\~n}a Hernandez} I.,  {D'Emilio} V.,
  {Morisaki} S.,  2023, \mn@doi [arXiv e-prints] {10.48550/arXiv.2306.00948},
  \href {https://ui.adsabs.harvard.edu/abs/2023arXiv230600948B} {p.
  arXiv:2306.00948}

\bibitem[\protect\citeauthoryear{{Bochenek}, {Ravi}, {Belov}, {Hallinan},
  {Kocz}, {Kulkarni}  \& {McKenna}}{{Bochenek} et~al.}{2020}]{frb_magnetar1}
{Bochenek} C.~D.,  {Ravi} V.,  {Belov} K.~V.,  {Hallinan} G.,  {Kocz} J.,
  {Kulkarni} S.~R.,   {McKenna} D.~L.,  2020, \mn@doi [\nat]
  {10.1038/s41586-020-2872-x}, \href
  {https://ui.adsabs.harvard.edu/abs/2020Natur.587...59B} {587, 59}

\bibitem[\protect\citeauthoryear{{CHIME/FRB Collaboration} et~al.,}{{CHIME/FRB
  Collaboration} et~al.}{2018}]{CHIME}
{CHIME/FRB Collaboration} et~al., 2018, \mn@doi [\apj]
  {10.3847/1538-4357/aad188}, \href
  {https://ui.adsabs.harvard.edu/abs/2018ApJ...863...48C} {863, 48}

\bibitem[\protect\citeauthoryear{{CHIME/FRB Collaboration} et~al.,}{{CHIME/FRB
  Collaboration} et~al.}{2019a}]{repeating_frb1}
{CHIME/FRB Collaboration} et~al., 2019a, \mn@doi [\nat]
  {10.1038/s41586-018-0867-7}, \href
  {https://ui.adsabs.harvard.edu/abs/2019Natur.566..230C} {566, 230}

\bibitem[\protect\citeauthoryear{{CHIME/FRB Collaboration} et~al.,}{{CHIME/FRB
  Collaboration} et~al.}{2019b}]{CHIME_frb_catalog_old}
{CHIME/FRB Collaboration} et~al., 2019b, \mn@doi [\apjl]
  {10.3847/2041-8213/ab4a80}, \href
  {https://ui.adsabs.harvard.edu/abs/2019ApJ...885L..24C} {885, L24}

\bibitem[\protect\citeauthoryear{{CHIME/FRB Collaboration} et~al.,}{{CHIME/FRB
  Collaboration} et~al.}{2019c}]{repeating_frb2}
{CHIME/FRB Collaboration} et~al., 2019c, \mn@doi [\apjl]
  {10.3847/2041-8213/ab4a80}, \href
  {https://ui.adsabs.harvard.edu/abs/2019ApJ...885L..24C} {885, L24}

\bibitem[\protect\citeauthoryear{{CHIME/FRB Collaboration} et~al.,}{{CHIME/FRB
  Collaboration} et~al.}{2021}]{CHIME_frb_catalog1}
{CHIME/FRB Collaboration} et~al., 2021, \mn@doi [\apjs]
  {10.3847/1538-4365/ac33ab10.48550/arXiv.2106.04352}, \href
  {https://ui.adsabs.harvard.edu/abs/2021ApJS..257...59C} {257, 59}

\bibitem[\protect\citeauthoryear{{Caleb} et~al.,}{{Caleb}
  et~al.}{2017}]{frb_caleb_et_al}
{Caleb} M.,  et~al., 2017, \mn@doi [\mnras] {10.1093/mnras/stx638}, \href
  {https://ui.adsabs.harvard.edu/abs/2017MNRAS.468.3746C} {468, 3746}

\bibitem[\protect\citeauthoryear{Cao, Li  \& Wang}{Cao
  et~al.}{2014}]{microlensing_cao_et_al}
Cao Z.,  Li L.-F.,   Wang Y.,  2014, \mn@doi [Phys. Rev. D]
  {10.1103/PhysRevD.90.062003}, 90, 062003

\bibitem[\protect\citeauthoryear{{Cao}, {Yu}  \& {Zhou}}{{Cao}
  et~al.}{2018}]{frb_beaming}
{Cao} X.-F.,  {Yu} Y.-W.,   {Zhou} X.,  2018, \mn@doi [\apj]
  {10.3847/1538-4357/aabadd}, \href
  {https://ui.adsabs.harvard.edu/abs/2018ApJ...858...89C} {858, 89}

\bibitem[\protect\citeauthoryear{{Champion} et~al.,}{{Champion}
  et~al.}{2016}]{frbs_champion_et_al}
{Champion} D.~J.,  et~al., 2016, \mn@doi [\mnras] {10.1093/mnrasl/slw069},
  \href {https://ui.adsabs.harvard.edu/abs/2016MNRAS.460L..30C} {460, L30}

\bibitem[\protect\citeauthoryear{{Chassande-Mottin}, {Leyde}, {Mastrogiovanni}
  \& {Steer}}{{Chassande-Mottin} et~al.}{2019}]{distance_uncertainties_GW}
{Chassande-Mottin} E.,  {Leyde} K.,  {Mastrogiovanni} S.,   {Steer} D.~A.,
  2019, \mn@doi [\prd] {10.1103/PhysRevD.100.083514}, \href
  {https://ui.adsabs.harvard.edu/abs/2019PhRvD.100h3514C} {100, 083514}

\bibitem[\protect\citeauthoryear{{Cheung}, {Gais}, {Hannuksela}  \&
  {Li}}{{Cheung} et~al.}{2021}]{Cheung_et_al_microlensing}
{Cheung} M. H.~Y.,  {Gais} J.,  {Hannuksela} O.~A.,   {Li} T. G.~F.,  2021,
  \mn@doi [\mnras] {10.1093/mnras/stab579}, \href
  {https://ui.adsabs.harvard.edu/abs/2021MNRAS.503.3326C} {503, 3326}

\bibitem[\protect\citeauthoryear{{Choi}, {Park}  \& {Vogeley}}{{Choi}
  et~al.}{2007}]{sdss_old}
{Choi} Y.-Y.,  {Park} C.,   {Vogeley} M.~S.,  2007, \mn@doi [\apj]
  {10.1086/511060}, \href
  {https://ui.adsabs.harvard.edu/abs/2007ApJ...658..884C} {658, 884}

\bibitem[\protect\citeauthoryear{{Christian}, {Vitale}  \& {Loeb}}{{Christian}
  et~al.}{2018}]{microlensing_christian_et_al}
{Christian} P.,  {Vitale} S.,   {Loeb} A.,  2018, \mn@doi [\prd]
  {10.1103/PhysRevD.98.103022}, \href
  {https://ui.adsabs.harvard.edu/abs/2018PhRvD..98j3022C} {98, 103022}

\bibitem[\protect\citeauthoryear{{Collett}}{{Collett}}{2015}]{sdss}
{Collett} T.~E.,  2015, \mn@doi [\apj] {10.1088/0004-637X/811/1/20}, \href
  {https://ui.adsabs.harvard.edu/abs/2015ApJ...811...20C} {811, 20}

\bibitem[\protect\citeauthoryear{{Collett} \& {Bacon}}{{Collett} \&
  {Bacon}}{2017}]{GW_speed_with_lensing1}
{Collett} T.~E.,  {Bacon} D.,  2017, \mn@doi [\prl]
  {10.1103/PhysRevLett.118.091101}, \href
  {https://ui.adsabs.harvard.edu/abs/2017PhRvL.118i1101C} {118, 091101}

\bibitem[\protect\citeauthoryear{{Cordes} \& {Chatterjee}}{{Cordes} \&
  {Chatterjee}}{2019}]{lensing_frb_review}
{Cordes} J.~M.,  {Chatterjee} S.,  2019, \mn@doi [\araa]
  {10.1146/annurev-astro-091918-104501}, \href
  {https://ui.adsabs.harvard.edu/abs/2019ARA&A..57..417C} {57, 417}

\bibitem[\protect\citeauthoryear{{Cordes}, {Wasserman}, {Hessels}, {Lazio},
  {Chatterjee}  \& {Wharton}}{{Cordes} et~al.}{2017}]{frb_plasma_lensing1}
{Cordes} J.~M.,  {Wasserman} I.,  {Hessels} J.~W.~T.,  {Lazio} T.~J.~W.,
  {Chatterjee} S.,   {Wharton} R.~S.,  2017, \mn@doi [\apj]
  {10.3847/1538-4357/aa74da}, \href
  {https://ui.adsabs.harvard.edu/abs/2017ApJ...842...35C} {842, 35}

\bibitem[\protect\citeauthoryear{{Dai} \& {Lu}}{{Dai} \&
  {Lu}}{2017}]{Dai_and_Lu_frb_lensing}
{Dai} L.,  {Lu} W.,  2017, \mn@doi [\apj] {10.3847/1538-4357/aa8873}, \href
  {https://ui.adsabs.harvard.edu/abs/2017ApJ...847...19D} {847, 19}

\bibitem[\protect\citeauthoryear{{Dai} \& {Venumadhav}}{{Dai} \&
  {Venumadhav}}{2017}]{Dai_Teja_strong_lensing}
{Dai} L.,  {Venumadhav} T.,  2017, \mn@doi [arXiv e-prints]
  {10.48550/arXiv.1702.04724}, \href
  {https://ui.adsabs.harvard.edu/abs/2017arXiv170204724D} {p. arXiv:1702.04724}

\bibitem[\protect\citeauthoryear{Dai, Li, Zackay, Mao  \& Lu}{Dai
  et~al.}{2018}]{Dai_et_al_microlensing}
Dai L.,  Li S.-S.,  Zackay B.,  Mao S.,   Lu Y.,  2018, \mn@doi [Phys. Rev. D]
  {10.1103/PhysRevD.98.104029}, 98, 104029

\bibitem[\protect\citeauthoryear{{Dai}, {Zackay}, {Venumadhav}, {Roulet}  \&
  {Zaldarriaga}}{{Dai} et~al.}{2020}]{Dai_et_al_lensing_search_O2}
{Dai} L.,  {Zackay} B.,  {Venumadhav} T.,  {Roulet} J.,   {Zaldarriaga} M.,
  2020, \mn@doi [arXiv e-prints] {10.48550/arXiv.2007.12709}, \href
  {https://ui.adsabs.harvard.edu/abs/2020arXiv200712709D} {p. arXiv:2007.12709}

\bibitem[\protect\citeauthoryear{{Deguchi} \& {Watson}}{{Deguchi} \&
  {Watson}}{1986}]{lensing_gw_2}
{Deguchi} S.,  {Watson} W.~D.,  1986, \mn@doi [\apj] {10.1086/164389}, \href
  {https://ui.adsabs.harvard.edu/abs/1986ApJ...307...30D} {307, 30}

\bibitem[\protect\citeauthoryear{{Diego}, {Hannuksela}, {Kelly}, {Pagano},
  {Broadhurst}, {Kim}, {Li}  \& {Smoot}}{{Diego}
  et~al.}{2019}]{microlensing_diego_et_al}
{Diego} J.~M.,  {Hannuksela} O.~A.,  {Kelly} P.~L.,  {Pagano} G.,  {Broadhurst}
  T.,  {Kim} K.,  {Li} T.~G.~F.,   {Smoot} G.~F.,  2019, \mn@doi [\aap]
  {10.1051/0004-6361/201935490}, \href
  {https://ui.adsabs.harvard.edu/abs/2019A&A...627A.130D} {627, A130}

\bibitem[\protect\citeauthoryear{{Ezquiaga}, {Holz}, {Hu}, {Lagos}  \&
  {Wald}}{{Ezquiaga} et~al.}{2021}]{Jose_et_al_strong_lensing}
{Ezquiaga} J.~M.,  {Holz} D.~E.,  {Hu} W.,  {Lagos} M.,   {Wald} R.~M.,  2021,
  \mn@doi [\prd] {10.1103/PhysRevD.103.064047}, \href
  {https://ui.adsabs.harvard.edu/abs/2021PhRvD.103f4047E} {103, 064047}

\bibitem[\protect\citeauthoryear{{Fairhurst}}{{Fairhurst}}{2009}]{Fairhurst_2009}
{Fairhurst} S.,  2009, \mn@doi [New Journal of Physics]
  {10.1088/1367-2630/11/12/123006}, \href
  {https://ui.adsabs.harvard.edu/abs/2009NJPh...11l3006F} {11, 123006}

\bibitem[\protect\citeauthoryear{{Falcke} \& {Rezzolla}}{{Falcke} \&
  {Rezzolla}}{2014}]{frb_supramassive_NS1}
{Falcke} H.,  {Rezzolla} L.,  2014, \mn@doi [\aap]
  {10.1051/0004-6361/20132199610.48550/arXiv.1307.1409}, \href
  {https://ui.adsabs.harvard.edu/abs/2014A&A...562A.137F} {562, A137}

\bibitem[\protect\citeauthoryear{{Fan}, {Liao}, {Biesiada},
  {Pi{\'o}rkowska-Kurpas}  \& {Zhu}}{{Fan} et~al.}{2017}]{GW_speed_lensing2}
{Fan} X.-L.,  {Liao} K.,  {Biesiada} M.,  {Pi{\'o}rkowska-Kurpas} A.,   {Zhu}
  Z.-H.,  2017, \mn@doi [\prl] {10.1103/PhysRevLett.118.091102}, \href
  {https://ui.adsabs.harvard.edu/abs/2017PhRvL.118i1102F} {118, 091102}

\bibitem[\protect\citeauthoryear{{Hannuksela}, {Haris}, {Ng}, {Kumar}, {Mehta},
  {Keitel}, {Li}  \& {Ajith}}{{Hannuksela} et~al.}{2019}]{lensing_otto}
{Hannuksela} O.~A.,  {Haris} K.,  {Ng} K.~K.~Y.,  {Kumar} S.,  {Mehta} A.~K.,
  {Keitel} D.,  {Li} T.~G.~F.,   {Ajith} P.,  2019, \mn@doi [\apjl]
  {10.3847/2041-8213/ab0c0f}, \href
  {https://ui.adsabs.harvard.edu/abs/2019ApJ...874L...2H} {874, L2}

\bibitem[\protect\citeauthoryear{{Haris}, {Mehta}, {Kumar}, {Venumadhav}  \&
  {Ajith}}{{Haris} et~al.}{2018}]{haris_et_al}
{Haris} K.,  {Mehta} A.~K.,  {Kumar} S.,  {Venumadhav} T.,   {Ajith} P.,  2018,
  arXiv e-prints, \href {https://ui.adsabs.harvard.edu/abs/2018arXiv180707062H}
  {p. arXiv:1807.07062}

\bibitem[\protect\citeauthoryear{{Hilbert}, {White}, {Hartlap}  \&
  {Schneider}}{{Hilbert} et~al.}{2007}]{multiple_cosmo_lensing}
{Hilbert} S.,  {White} S. D.~M.,  {Hartlap} J.,   {Schneider} P.,  2007,
  \mn@doi [\mnras] {10.1111/j.1365-2966.2007.12391.x}, \href
  {https://ui.adsabs.harvard.edu/abs/2007MNRAS.382..121H} {382, 121}

\bibitem[\protect\citeauthoryear{Hild}{Hild}{2012}]{ET_PSD}
Hild S.,  2012, \mn@doi [Classical and Quantum Gravity]
  {10.1088/0264-9381/29/12/124006}, 29, 124006

\bibitem[\protect\citeauthoryear{{James}, {Prochaska}, {Macquart},
  {North-Hickey}, {Bannister}  \& {Dunning}}{{James}
  et~al.}{2022}]{z_DM_distribution_frbs}
{James} C.~W.,  {Prochaska} J.~X.,  {Macquart} J.~P.,  {North-Hickey} F.~O.,
  {Bannister} K.~W.,   {Dunning} A.,  2022, \mn@doi [\mnras]
  {10.1093/mnras/stab3051}, \href
  {https://ui.adsabs.harvard.edu/abs/2022MNRAS.509.4775J} {509, 4775}

\bibitem[\protect\citeauthoryear{KAGRA, Collaboration  \& Collaboration}{KAGRA
  et~al.}{2019}]{observer_summary}
KAGRA C.,  Collaboration L.~S.,   Collaboration V.,  2019, Advanced LIGO,
  Advanced Virgo and KAGRA observing run plans, \url
  {https://dcc.ligo.org/public/0161/P1900218/002/SummaryForObservers.pdf}

\bibitem[\protect\citeauthoryear{{Kader} et~al.,}{{Kader}
  et~al.}{2022}]{microlensed_frbs_zarif}
{Kader} Z.,  et~al., 2022, \mn@doi [\prd] {10.1103/PhysRevD.106.043016}, \href
  {https://ui.adsabs.harvard.edu/abs/2022PhRvD.106d3016K} {106, 043016}

\bibitem[\protect\citeauthoryear{{Kormann}, {Schneider}  \&
  {Bartelmann}}{{Kormann} et~al.}{1994}]{kormann_et_al}
{Kormann} R.,  {Schneider} P.,   {Bartelmann} M.,  1994, \aap, \href
  {https://ui.adsabs.harvard.edu/abs/1994A&A...284..285K} {284, 285}

\bibitem[\protect\citeauthoryear{{Kumar} \& {Beniamini}}{{Kumar} \&
  {Beniamini}}{2022}]{frb_plasma_lensing2}
{Kumar} P.,  {Beniamini} P.,  2022, \mn@doi [arXiv e-prints]
  {10.48550/arXiv.2208.03332}, \href
  {https://ui.adsabs.harvard.edu/abs/2022arXiv220803332K} {p. arXiv:2208.03332}

\bibitem[\protect\citeauthoryear{Kumar \& Linder}{Kumar \&
  Linder}{2019}]{distance_DM_frb}
Kumar P.,  Linder E.~V.,  2019, \mn@doi [Phys. Rev. D]
  {10.1103/PhysRevD.100.083533}, 100, 083533

\bibitem[\protect\citeauthoryear{{Kumar} et~al.,}{{Kumar}
  et~al.}{2019}]{repeating_frb3}
{Kumar} P.,  et~al., 2019, \mn@doi [\apjl] {10.3847/2041-8213/ab5b08}, \href
  {https://ui.adsabs.harvard.edu/abs/2019ApJ...887L..30K} {887, L30}

\bibitem[\protect\citeauthoryear{{LIGO Scientific Collaboration}}{{LIGO
  Scientific Collaboration}}{2015}]{Voyager_PSD}
{LIGO Scientific Collaboration} 2015, Instrument Science White Paper, \url
  {https://dcc.ligo.org/public/0120/T1500290/002/T1500290.pdf}

\bibitem[\protect\citeauthoryear{{LIGO Scientific Collaboration}}{{LIGO
  Scientific Collaboration}}{2018}]{instrument_science}
{LIGO Scientific Collaboration} 2018, Instrument Science White Paper 2018, \url
  {https://dcc.ligo.org/public/0151/T1800133/004/T1800133-instrument-science-white-v4.pdf}

\bibitem[\protect\citeauthoryear{Lai, Hannuksela, Herrera-Mart\'{\i}n, Diego,
  Broadhurst  \& Li}{Lai et~al.}{2018}]{microlensing_lai_et_al}
Lai K.-H.,  Hannuksela O.~A.,  Herrera-Mart\'{\i}n A.,  Diego J.~M.,
  Broadhurst T.,   Li T. G.~F.,  2018, \mn@doi [Phys. Rev. D]
  {10.1103/PhysRevD.98.083005}, 98, 083005

\bibitem[\protect\citeauthoryear{Leung et~al.,}{Leung
  et~al.}{2022}]{Leung_et_al_frb_lensing}
Leung C.,  et~al., 2022, \mn@doi [Phys. Rev. D] {10.1103/PhysRevD.106.043017},
  106, 043017

\bibitem[\protect\citeauthoryear{{Li} \& {Li}}{{Li} \&
  {Li}}{2014}]{Li_and_Li_frb_lensing}
{Li} C.,  {Li} L.,  2014, \mn@doi [Science China Physics, Mechanics, and
  Astronomy] {10.1007/s11433-014-5465-6}, \href
  {https://ui.adsabs.harvard.edu/abs/2014SCPMA..57.1390L} {57, 1390}

\bibitem[\protect\citeauthoryear{{Li}, {Mao}, {Zhao}  \& {Lu}}{{Li}
  et~al.}{2018}]{Li_et_al_strong_lensing}
{Li} S.-S.,  {Mao} S.,  {Zhao} Y.,   {Lu} Y.,  2018, \mn@doi [\mnras]
  {10.1093/mnras/sty411}, \href
  {https://ui.adsabs.harvard.edu/abs/2018MNRAS.476.2220L} {476, 2220}

\bibitem[\protect\citeauthoryear{{Li}, {Lo}, {Sachdev}, {Chan}, {Lin}, {Li}  \&
  {Weinstein}}{{Li} et~al.}{2019}]{Alvin_et_al_lensing_search_LV}
{Li} A. K.~Y.,  {Lo} R. K.~L.,  {Sachdev} S.,  {Chan} C.~L.,  {Lin} E.~T.,
  {Li} T. G.~F.,   {Weinstein} A.~J.,  2019, \mn@doi [arXiv e-prints]
  {10.48550/arXiv.1904.06020}, \href
  {https://ui.adsabs.harvard.edu/abs/2019arXiv190406020L} {p. arXiv:1904.06020}

\bibitem[\protect\citeauthoryear{{Lin} et~al.,}{{Lin} et~al.}{2022}]{BURSTT}
{Lin} H.-H.,  et~al., 2022, \mn@doi [\pasp] {10.1088/1538-3873/ac8f71}, \href
  {https://ui.adsabs.harvard.edu/abs/2022PASP..134i4106L} {134, 094106}

\bibitem[\protect\citeauthoryear{{Liu}, {Maga{\~n}a Hernandez}  \&
  {Creighton}}{{Liu} et~al.}{2021}]{Liu_et_al_lensing_search_O2}
{Liu} X.,  {Maga{\~n}a Hernandez} I.,   {Creighton} J.,  2021, \mn@doi [\apj]
  {10.3847/1538-4357/abd7eb}, \href
  {https://ui.adsabs.harvard.edu/abs/2021ApJ...908...97L} {908, 97}

\bibitem[\protect\citeauthoryear{Lorimer, Bailes, McLaughlin, Narkevic  \&
  Crawford}{Lorimer et~al.}{2007}]{lorimer_FRB}
Lorimer D.~R.,  Bailes M.,  McLaughlin M.~A.,  Narkevic D.~J.,   Crawford F.,
  2007, Science, 318, 777

\bibitem[\protect\citeauthoryear{{Madau} \& {Dickinson}}{{Madau} \&
  {Dickinson}}{2014}]{madau_dickinson}
{Madau} P.,  {Dickinson} M.,  2014, \mn@doi [\araa]
  {10.1146/annurev-astro-081811-125615}, \href
  {https://ui.adsabs.harvard.edu/abs/2014ARA&A..52..415M} {52, 415}

\bibitem[\protect\citeauthoryear{{Magare}, {Kapadia}, {More}, {Singh}, {Ajith}
  \& {Ramprakash}}{{Magare} et~al.}{2023}]{magare2023}
{Magare} S.,  {Kapadia} S.~J.,  {More} A.,  {Singh} M.~K.,  {Ajith} P.,
  {Ramprakash} A.~N.,  2023, \mn@doi [arXiv e-prints]
  {10.48550/arXiv.2302.02916}, \href
  {https://ui.adsabs.harvard.edu/abs/2023arXiv230202916M} {p. arXiv:2302.02916}

\bibitem[\protect\citeauthoryear{{Masui} et~al.,}{{Masui}
  et~al.}{2015}]{frb_masui_et_al}
{Masui} K.,  et~al., 2015, \mn@doi [\nat] {10.1038/nature15769}, \href
  {https://ui.adsabs.harvard.edu/abs/2015Natur.528..523M} {528, 523}

\bibitem[\protect\citeauthoryear{{McIsaac}, {Keitel}, {Collett}, {Harry},
  {Mozzon}, {Edy}  \& {Bacon}}{{McIsaac}
  et~al.}{2020}]{Connor_et_al_lensing_search_O1O2}
{McIsaac} C.,  {Keitel} D.,  {Collett} T.,  {Harry} I.,  {Mozzon} S.,  {Edy}
  O.,   {Bacon} D.,  2020, \mn@doi [\prd] {10.1103/PhysRevD.102.084031}, \href
  {https://ui.adsabs.harvard.edu/abs/2020PhRvD.102h4031M} {102, 084031}

\bibitem[\protect\citeauthoryear{{Mishra}, {Meena}, {More}, {Bose}  \&
  {Bagla}}{{Mishra} et~al.}{2021}]{Mishra_et_al_microlensing}
{Mishra} A.,  {Meena} A.~K.,  {More} A.,  {Bose} S.,   {Bagla} J.~S.,  2021,
  \mn@doi [\mnras] {10.1093/mnras/stab2875}, \href
  {https://ui.adsabs.harvard.edu/abs/2021MNRAS.508.4869M} {508, 4869}

\bibitem[\protect\citeauthoryear{{Moortgat} \& {Kuijpers}}{{Moortgat} \&
  {Kuijpers}}{2005}]{frb_BNS_circumbinary_plasma}
{Moortgat} J.,  {Kuijpers} J.,  2005, in 22nd Texas Symposium on Relativistic
  Astrophysics. pp 326--331 (\mn@eprint {arXiv} {gr-qc/0503074}),
  \mn@doi{10.48550/arXiv.gr-qc/0503074}

\bibitem[\protect\citeauthoryear{{Moroianu}, {Wen}, {James}, {Ai}, {Kovalam},
  {Panther}  \& {Zhang}}{{Moroianu} et~al.}{2022}]{FRB_BNS_association}
{Moroianu} A.,  {Wen} L.,  {James} C.~W.,  {Ai} S.,  {Kovalam} M.,  {Panther}
  F.,   {Zhang} B.,  2022, \mn@doi [arXiv e-prints]
  {10.48550/arXiv.2212.00201}, \href
  {https://ui.adsabs.harvard.edu/abs/2022arXiv221200201M} {p. arXiv:2212.00201}

\bibitem[\protect\citeauthoryear{{Mu{\~n}oz}, {Kovetz}, {Dai}  \&
  {Kamionkowski}}{{Mu{\~n}oz} et~al.}{2016}]{Munoz_et_al_frb_lensing}
{Mu{\~n}oz} J.~B.,  {Kovetz} E.~D.,  {Dai} L.,   {Kamionkowski} M.,  2016,
  \mn@doi [\prl] {10.1103/PhysRevLett.117.091301}, \href
  {https://ui.adsabs.harvard.edu/abs/2016PhRvL.117i1301M} {117, 091301}

\bibitem[\protect\citeauthoryear{Nakamura}{Nakamura}{1998}]{lensing_gw_4}
Nakamura T.~T.,  1998, \mn@doi [Phys. Rev. Lett.]
  {10.1103/PhysRevLett.80.1138}, 80, 1138

\bibitem[\protect\citeauthoryear{{Ng}, {Wong}, {Broadhurst}  \& {Li}}{{Ng}
  et~al.}{2018}]{Ng_et_al_strong_lensing}
{Ng} K. K.~Y.,  {Wong} K. W.~K.,  {Broadhurst} T.,   {Li} T. G.~F.,  2018,
  \mn@doi [\prd] {10.1103/PhysRevD.97.023012}, \href
  {https://ui.adsabs.harvard.edu/abs/2018PhRvD..97b3012N} {97, 023012}

\bibitem[\protect\citeauthoryear{Nitz, Capano, Nielsen, Reyes, White, Brown  \&
  Krishnan}{Nitz et~al.}{2019}]{Nitz_OGC1}
Nitz A.~H.,  Capano C.,  Nielsen A.~B.,  Reyes S.,  White R.,  Brown D.~A.,
  Krishnan B.,  2019, \mn@doi [The Astrophysical Journal]
  {10.3847/1538-4357/ab0108}, 872, 195

\bibitem[\protect\citeauthoryear{Nitz et~al.,}{Nitz et~al.}{2020}]{Nitz_OGC2}
Nitz A.~H.,  et~al., 2020, \mn@doi [The Astrophysical Journal]
  {10.3847/1538-4357/ab733f}, 891, 123

\bibitem[\protect\citeauthoryear{{Nitz}, {Kumar}, {Wang}, {Kastha}, {Wu},
  {Sch{\"a}fer}, {Dhurkunde}  \& {Capano}}{{Nitz} et~al.}{2021a}]{Nitz_OGC4}
{Nitz} A.~H.,  {Kumar} S.,  {Wang} Y.-F.,  {Kastha} S.,  {Wu} S.,
  {Sch{\"a}fer} M.,  {Dhurkunde} R.,   {Capano} C.~D.,  2021a, arXiv e-prints,
  \href {https://ui.adsabs.harvard.edu/abs/2021arXiv211206878N} {p.
  arXiv:2112.06878}

\bibitem[\protect\citeauthoryear{Nitz, Capano, Kumar, Wang, Kastha, Schäfer,
  Dhurkunde  \& Cabero}{Nitz et~al.}{2021b}]{Nitz_OGC3}
Nitz A.~H.,  Capano C.~D.,  Kumar S.,  Wang Y.-F.,  Kastha S.,  Schäfer M.,
  Dhurkunde R.,   Cabero M.,  2021b, \mn@doi [The Astrophysical Journal]
  {10.3847/1538-4357/ac1c03}, 922, 76

\bibitem[\protect\citeauthoryear{{Oguri}}{{Oguri}}{2018}]{Oguri_et_al_strong_lensing}
{Oguri} M.,  2018, \mn@doi [\mnras] {10.1093/mnras/sty2145}, \href
  {https://ui.adsabs.harvard.edu/abs/2018MNRAS.480.3842O} {480, 3842}

\bibitem[\protect\citeauthoryear{{Ohanian}}{{Ohanian}}{1974}]{lensing_gw_1}
{Ohanian} H.~C.,  1974, \mn@doi [International Journal of Theoretical Physics]
  {10.1007/BF01810927}, \href
  {https://ui.adsabs.harvard.edu/abs/1974IJTP....9..425O} {9, 425}

\bibitem[\protect\citeauthoryear{{Pang}, {Hannuksela}, {Dietrich}, {Pagano}  \&
  {Harry}}{{Pang} et~al.}{2020}]{Pang_et_al_lensing_search_GW}
{Pang} P. T.~H.,  {Hannuksela} O.~A.,  {Dietrich} T.,  {Pagano} G.,   {Harry}
  I.~W.,  2020, \mn@doi [\mnras] {10.1093/mnras/staa1430}, \href
  {https://ui.adsabs.harvard.edu/abs/2020MNRAS.495.3740P} {495, 3740}

\bibitem[\protect\citeauthoryear{{Petroff} et~al.,}{{Petroff}
  et~al.}{2016}]{frb_catalog_petroff_et_al}
{Petroff} E.,  et~al., 2016, \mn@doi [\pasa] {10.1017/pasa.2016.35}, \href
  {https://ui.adsabs.harvard.edu/abs/2016PASA...33...45P} {33, e045}

\bibitem[\protect\citeauthoryear{{Pitkin}, {Reid}, {Rowan}  \&
  {Hough}}{{Pitkin} et~al.}{2011}]{AdvLIGO_Virgo}
{Pitkin} M.,  {Reid} S.,  {Rowan} S.,   {Hough} J.,  2011, \mn@doi [Living
  Reviews in Relativity] {10.12942/lrr-2011-5}, \href
  {https://ui.adsabs.harvard.edu/abs/2011LRR....14....5P} {14, 5}

\bibitem[\protect\citeauthoryear{{Platts}, {Weltman}, {Walters}, {Tendulkar},
  {Gordin}  \& {Kandhai}}{{Platts} et~al.}{2019}]{frb_theories_living}
{Platts} E.,  {Weltman} A.,  {Walters} A.,  {Tendulkar} S.~P.,  {Gordin}
  J.~E.~B.,   {Kandhai} S.,  2019, \mn@doi [\physrep]
  {10.1016/j.physrep.2019.06.00310.48550/arXiv.1810.05836}, \href
  {https://ui.adsabs.harvard.edu/abs/2019PhR...821....1P} {821, 1}

\bibitem[\protect\citeauthoryear{{Pshirkov} \& {Postnov}}{{Pshirkov} \&
  {Postnov}}{2010}]{frb_precursor_BNS}
{Pshirkov} M.~S.,  {Postnov} K.~A.,  2010, \mn@doi [\apss]
  {10.1007/s10509-010-0395-x10.48550/arXiv.1004.5115}, \href
  {https://ui.adsabs.harvard.edu/abs/2010Ap&SS.330...13P} {330, 13}

\bibitem[\protect\citeauthoryear{Punturo et~al.}{Punturo et~al.}{2010}]{ET}
Punturo M.,  et~al., 2010, \mn@doi [Classical and Quantum Gravity]
  {10.1088/0264-9381/27/19/194002}, 27, 194002

\bibitem[\protect\citeauthoryear{{Ravi} \& {Lasky}}{{Ravi} \&
  {Lasky}}{2014}]{NS_collapse_times}
{Ravi} V.,  {Lasky} P.~D.,  2014, \mn@doi [\mnras]
  {10.1093/mnras/stu72010.48550/arXiv.1403.6327}, \href
  {https://ui.adsabs.harvard.edu/abs/2014MNRAS.441.2433R} {441, 2433}

\bibitem[\protect\citeauthoryear{{Reitze} et~al.}{{Reitze} et~al.}{2019}]{CE}
{Reitze} D.,  et~al., 2019, in \baas. p.~35 (\mn@eprint {arXiv} {1907.04833})

\bibitem[\protect\citeauthoryear{{Sagiv} \& {Waxman}}{{Sagiv} \&
  {Waxman}}{2002}]{Sagiv_GRB_shock}
{Sagiv} A.,  {Waxman} E.,  2002, \mn@doi [\apj] {10.1086/340948}, \href
  {https://ui.adsabs.harvard.edu/abs/2002ApJ...574..861S} {574, 861}

\bibitem[\protect\citeauthoryear{{Saleem} et~al.,}{{Saleem}
  et~al.}{2021}]{SaleemLIGOIndia}
{Saleem} M.,  et~al., 2021, arXiv e-prints, \href
  {https://ui.adsabs.harvard.edu/abs/2021arXiv210501716S} {p. arXiv:2105.01716}

\bibitem[\protect\citeauthoryear{{Shannon} et~al.,}{{Shannon}
  et~al.}{2018}]{frb_shannon_et_al}
{Shannon} R.~M.,  et~al., 2018, \mn@doi [\nat] {10.1038/s41586-018-0588-y},
  \href {https://ui.adsabs.harvard.edu/abs/2018Natur.562..386S} {562, 386}

\bibitem[\protect\citeauthoryear{{Shin} et~al.,}{{Shin}
  et~al.}{2023}]{distance_uncertainties_FRB}
{Shin} K.,  et~al., 2023, \mn@doi [\apj] {10.3847/1538-4357/acaf06}, \href
  {https://ui.adsabs.harvard.edu/abs/2023ApJ...944..105S} {944, 105}

\bibitem[\protect\citeauthoryear{Singer \& Price}{Singer \&
  Price}{2016}]{Singer_2016}
Singer L.~P.,  Price L.~R.,  2016, \mn@doi [Phys. Rev. D]
  {10.1103/PhysRevD.93.024013}, 93, 024013

\bibitem[\protect\citeauthoryear{Singer et~al.,}{Singer
  et~al.}{2014}]{Singer_2014}
Singer L.~P.,  et~al., 2014, \mn@doi [The Astrophysical Journal]
  {10.1088/0004-637X/795/2/105}, 795, 105

\bibitem[\protect\citeauthoryear{{Spitler} et~al.,}{{Spitler}
  et~al.}{2014}]{frb_arecibo}
{Spitler} L.~G.,  et~al., 2014, \mn@doi [\apj] {10.1088/0004-637X/790/2/101},
  \href {https://ui.adsabs.harvard.edu/abs/2014ApJ...790..101S} {790, 101}

\bibitem[\protect\citeauthoryear{{Takahashi} \& {Nakamura}}{{Takahashi} \&
  {Nakamura}}{2003}]{microlensing}
{Takahashi} R.,  {Nakamura} T.,  2003, \mn@doi [\apj] {10.1086/377430}, \href
  {https://ui.adsabs.harvard.edu/abs/2003ApJ...595.1039T} {595, 1039}

\bibitem[\protect\citeauthoryear{{Takahashi}, {Oguri}, {Sato}  \&
  {Hamana}}{{Takahashi} et~al.}{2011}]{cosmo_lensing2}
{Takahashi} R.,  {Oguri} M.,  {Sato} M.,   {Hamana} T.,  2011, \mn@doi [\apj]
  {10.1088/0004-637X/742/1/15}, \href
  {https://ui.adsabs.harvard.edu/abs/2011ApJ...742...15T} {742, 15}

\bibitem[\protect\citeauthoryear{{The LIGO Scientific Collaboration}, {the
  Virgo Collaboration}, {the KAGRA Collaboration}, {the CHIME/FRB
  Collaboration}, {:}, {Abbott}  et~al.}{{The LIGO Scientific Collaboration}
  et~al.}{2022}]{LVK_CBC_FRB}
{The LIGO Scientific Collaboration} {the Virgo Collaboration} {the KAGRA
  Collaboration} {the CHIME/FRB Collaboration} {:} {Abbott} R.,   et~al., 2022,
  arXiv e-prints, \href {https://ui.adsabs.harvard.edu/abs/2022arXiv220312038T}
  {p. arXiv:2203.12038}

\bibitem[\protect\citeauthoryear{{Thornton} et~al.,}{{Thornton}
  et~al.}{2013}]{frb_catalog_thornton_et_al}
{Thornton} D.,  et~al., 2013, \mn@doi [Science] {10.1126/science.1236789},
  \href {https://ui.adsabs.harvard.edu/abs/2013Sci...341...53T} {341, 53}

\bibitem[\protect\citeauthoryear{{Torchinsky}, {Broderick}, {Gunst}, {Faulkner}
   \& {van Cappellen}}{{Torchinsky} et~al.}{2016}]{SKA}
{Torchinsky} S.~A.,  {Broderick} J.~W.,  {Gunst} A.,  {Faulkner} A.~J.,   {van
  Cappellen} W.,  2016, \mn@doi [arXiv e-prints] {10.48550/arXiv.1610.00683},
  \href {https://ui.adsabs.harvard.edu/abs/2016arXiv161000683T} {p.
  arXiv:1610.00683}

\bibitem[\protect\citeauthoryear{{Totani}}{{Totani}}{2013}]{frb_BNS_merger}
{Totani} T.,  2013, \mn@doi [\pasj]
  {10.1093/pasj/65.5.L1210.48550/arXiv.1307.4985}, \href
  {https://ui.adsabs.harvard.edu/abs/2013PASJ...65L..12T} {65, L12}

\bibitem[\protect\citeauthoryear{{Usov} \& {Katz}}{{Usov} \&
  {Katz}}{2000}]{Usov_GRB_shock}
{Usov} V.~V.,  {Katz} J.~I.,  2000, \mn@doi [\aap]
  {10.48550/arXiv.astro-ph/0002278}, \href
  {https://ui.adsabs.harvard.edu/abs/2000A&A...364..655U} {364, 655}

\bibitem[\protect\citeauthoryear{{Venumadhav}, {Zackay}, {Roulet}, {Dai}  \&
  {Zaldarriaga}}{{Venumadhav} et~al.}{2019}]{ias-3}
{Venumadhav} T.,  {Zackay} B.,  {Roulet} J.,  {Dai} L.,   {Zaldarriaga} M.,
  2019, \mn@doi [\prd] {10.1103/PhysRevD.100.023011}, \href
  {https://ui.adsabs.harvard.edu/abs/2019PhRvD.100b3011V} {100, 023011}

\bibitem[\protect\citeauthoryear{{Venumadhav}, {Zackay}, {Roulet}, {Dai}  \&
  {Zaldarriaga}}{{Venumadhav} et~al.}{2020}]{ias-2}
{Venumadhav} T.,  {Zackay} B.,  {Roulet} J.,  {Dai} L.,   {Zaldarriaga} M.,
  2020, \mn@doi [\prd] {10.1103/PhysRevD.101.083030}, \href
  {https://ui.adsabs.harvard.edu/abs/2020PhRvD.101h3030V} {101, 083030}

\bibitem[\protect\citeauthoryear{{Vijaykumar}, {Saketh}, {Kumar}, {Ajith}  \&
  {Choudhury}}{{Vijaykumar} et~al.}{2020}]{Aditya_V_large_scale}
{Vijaykumar} A.,  {Saketh} M.~V.~S.,  {Kumar} S.,  {Ajith} P.,   {Choudhury}
  T.~R.,  2020, \mn@doi [arXiv e-prints] {10.48550/arXiv.2005.01111}, \href
  {https://ui.adsabs.harvard.edu/abs/2020arXiv200501111V} {p. arXiv:2005.01111}

\bibitem[\protect\citeauthoryear{Wang, Stebbins  \& Turner}{Wang
  et~al.}{1996}]{lensing_gw_3}
Wang Y.,  Stebbins A.,   Turner E.~L.,  1996, \mn@doi [Phys. Rev. Lett.]
  {10.1103/PhysRevLett.77.2875}, 77, 2875

\bibitem[\protect\citeauthoryear{{Wang}, {Yang}, {Wu}, {Dai}  \& {Wang}}{{Wang}
  et~al.}{2016}]{frb_BNS_inspiral}
{Wang} J.-S.,  {Yang} Y.-P.,  {Wu} X.-F.,  {Dai} Z.-G.,   {Wang} F.-Y.,  2016,
  \mn@doi [\apjl] {10.3847/2041-8205/822/1/L710.48550/arXiv.1603.02014}, \href
  {https://ui.adsabs.harvard.edu/abs/2016ApJ...822L...7W} {822, L7}

\bibitem[\protect\citeauthoryear{{Wang}, {Zhang}, {Dai}  \& {Cheng}}{{Wang}
  et~al.}{2022}]{frb_magnetar2}
{Wang} F.~Y.,  {Zhang} G.~Q.,  {Dai} Z.~G.,   {Cheng} K.~S.,  2022, \mn@doi
  [Nature Communications] {10.1038/s41467-022-31923-y}, \href
  {https://ui.adsabs.harvard.edu/abs/2022NatCo..13.4382W} {13, 4382}

\bibitem[\protect\citeauthoryear{{Xu} \& {Zhang}}{{Xu} \&
  {Zhang}}{2020}]{DM_Xu_Zhang}
{Xu} S.,  {Zhang} B.,  2020, \mn@doi [\apjl] {10.3847/2041-8213/aba760}, \href
  {https://ui.adsabs.harvard.edu/abs/2020ApJ...898L..48X} {898, L48}

\bibitem[\protect\citeauthoryear{Xu, Weinberg  \& Zhang}{Xu
  et~al.}{2021}]{DM_uncertainties_Xu_et_al}
Xu S.,  Weinberg D.~H.,   Zhang B.,  2021, \mn@doi [The Astrophysical Journal
  Letters] {10.3847/2041-8213/ac399c}, 922, L31

\bibitem[\protect\citeauthoryear{{Zackay}, {Dai}, {Venumadhav}, {Roulet}  \&
  {Zaldarriaga}}{{Zackay} et~al.}{2019a}]{ias-1}
{Zackay} B.,  {Dai} L.,  {Venumadhav} T.,  {Roulet} J.,   {Zaldarriaga} M.,
  2019a, arXiv e-prints, \href
  {https://ui.adsabs.harvard.edu/abs/2019arXiv191009528Z} {p. arXiv:1910.09528}

\bibitem[\protect\citeauthoryear{{Zackay}, {Venumadhav}, {Dai}, {Roulet}  \&
  {Zaldarriaga}}{{Zackay} et~al.}{2019b}]{ias-4}
{Zackay} B.,  {Venumadhav} T.,  {Dai} L.,  {Roulet} J.,   {Zaldarriaga} M.,
  2019b, \mn@doi [\prd] {10.1103/PhysRevD.100.023007}, \href
  {https://ui.adsabs.harvard.edu/abs/2019PhRvD.100b3007Z} {100, 023007}

\bibitem[\protect\citeauthoryear{{Zhang}}{{Zhang}}{2014}]{frb_supramassive_NS2}
{Zhang} B.,  2014, \mn@doi [\apjl]
  {10.1088/2041-8205/780/2/L2110.48550/arXiv.1310.4893}, \href
  {https://ui.adsabs.harvard.edu/abs/2014ApJ...780L..21Z} {780, L21}

\bibitem[\protect\citeauthoryear{{Zhang}}{{Zhang}}{2016}]{frb_charged_BH}
{Zhang} B.,  2016, \mn@doi [\apjl]
  {10.3847/2041-8205/827/2/L3110.48550/arXiv.1602.04542}, \href
  {https://ui.adsabs.harvard.edu/abs/2016ApJ...827L..31Z} {827, L31}

\bibitem[\protect\citeauthoryear{{Zhang}}{{Zhang}}{2019}]{frb_charged_BHNS}
{Zhang} B.,  2019, \mn@doi [\apjl] {10.3847/2041-8213/ab0ae8}, \href
  {https://ui.adsabs.harvard.edu/abs/2019ApJ...873L...9Z} {873, L9}

\bibitem[\protect\citeauthoryear{{Zhang}}{{Zhang}}{2020a}]{frb_review_zhang}
{Zhang} B.,  2020a, \mn@doi [\nat] {10.1038/s41586-020-2828-1}, \href
  {https://ui.adsabs.harvard.edu/abs/2020Natur.587...45Z} {587, 45}

\bibitem[\protect\citeauthoryear{{Zhang}}{{Zhang}}{2020b}]{frb_BNS_zhang}
{Zhang} B.,  2020b, \mn@doi [\apjl] {10.3847/2041-8213/ab7244}, \href
  {https://ui.adsabs.harvard.edu/abs/2020ApJ...890L..24Z} {890, L24}

\makeatother
\end{thebibliography}
